\PassOptionsToPackage{numbers,sort&compress}{natbib}
\documentclass[aps,pra,reprint,onecolumn,11pt,tightenlines,raggedbottom,notitlepage,nofootinbib,showkeys,amsmath,amssymb]{revtex4-2}

\usepackage[T1]{fontenc}
\usepackage{lmodern}
\usepackage{graphicx}
\usepackage{booktabs}
\usepackage{bm}
\usepackage{siunitx}
\usepackage{longtable}
\usepackage{placeins}
\usepackage{capt-of}
\usepackage[hidelinks]{hyperref}
\usepackage[capitalise,nameinlink]{cleveref}

\makeatletter
\def\pdfstartlink@attr{attr{/Border[0 0 0]/H/I}}
 \AtBeginDocument{\def\p@subsection{}\def\p@subsubsection{}}
\makeatother
\graphicspath{{figures/}}
\hypersetup{
  pdftitle={Secure Quantum Metasurfaces for Direct Quantum Key Distribution Measurements},
  pdfauthor={A Dada}
}
\renewcommand{\arraystretch}{1.14}
\setcitestyle{numbers,square,comma}
\renewcommand{\thesection}{\arabic{section}}
\renewcommand{\thesubsection}{\thesection.\arabic{subsection}}
\renewcommand{\thetable}{\arabic{table}}
\AtBeginDocument{%
}

\DeclareMathOperator{\Tr}{Tr}
\newcommand{\ket}[1]{\lvert #1\rangle}
\newcommand{\bra}[1]{\langle #1\rvert}
\newcommand{\proj}[1]{\ket{#1}\!\bra{#1}}
\newcommand{\identity}{\hat{\mathbb{I}}}
\newcommand{\binaryentropy}{h_2}
\newcommand{\loss}{\varnothing}

\begin{document}

\title{Secure Quantum Metasurfaces for Direct Quantum Key Distribution Measurements}
\author{Yan He}
\author{Adetunmise C. Dada}
\thanks{Correspondence: Adetunmise C. Dada (\href{mailto:adetunmise.dada@glasgow.ac.uk}{adetunmise.dada@glasgow.ac.uk})}
\affiliation{School of Physics and Astronomy, University of Glasgow, Glasgow, G12 8QQ, United Kingdom}
\date[]{}
\keywords{dielectric metasurfaces | quantum key distribution | BB84 | polarization measurement | flat optics}

\begin{abstract}
Quantum key distribution (QKD) can deliver information-theoretic security, but its receivers are still typically assembled from cascaded bulk optics.
Metasurfaces offer a radical reduction in this complexity.
Here, we make the metasurface function as the QKD measurement device itself.
We design wavelength-specific dielectric apertures at 780, 1550, 2000, and 10\,600~nm that interweave the linear $H/V$ and circular $R/L$ phase libraries at a 0.44--0.49$\lambda$ cell pitch, i.e., below half the operating wavelength, and map the complete passive-BB84 measurement directly onto four detector channels.
This subwavelength basis co-location prevents an ordinary propagating-wave aperture mask from isolating one basis without simultaneously attenuating or perturbing the other.
Simulated full-wave focal-plane intensities are converted directly into conditional Born probabilities using fitted, calibration-fixed detector regions.
The selected designs achieve probability fidelities of 97.96--99.08\%, four-port collected efficiencies of 17.43--40.94\%, and mean intrinsic quantum bit error rates of 1.42--3.96\%.
At 1550~nm, the selected design gives an asymptotic device-level yield of 0.151 secret bits per incident photon under an ideal single-photon model.
Six-state detector tomography gives positive both-basis security bounds for all six reconstructed receiver models, including their polarization-dependent loss, basis imbalance and crosstalk.
These results establish protocol-matched meta-optics as a route to compact, passive QKD receivers.
\end{abstract}
\maketitle

\section{Introduction}

Quantum key distribution (QKD) converts the disturbance associated with incompatible quantum measurements into an operational test of channel security. In polarization-encoded BB84, a receiver measures each photon in one of two mutually unbiased bases and retains matched-basis events for parameter estimation and key generation \cite{BennettBrassard1984,ShorPreskill2000,Scarani2009,Xu2020}. The protocol is compact, whereas a passive optical receiver generally requires a non-polarizing beam splitter, basis-rotation optics, two polarization beam splitters, several free-space paths, and four detector channels. This component chain potentially increases insertion loss, footprint, alignment burden, and the number of basis-dependent responses that must be characterized.

Dielectric metasurfaces can control phase, amplitude, polarization, and propagation direction within a subwavelength-thickness element \cite{YuCapasso2014,Kamali2018,Chen2020}. Their use in quantum photonics now extends from state generation and interference to compact polarization analysis and generalized measurements \cite{Wang2018,Solntsev2021,Shah2022,Lung2024,An2024}. However, the distinction between assisting a QKD system and performing its measurement is fundamental. Recent metasurface POVMs have targeted informationally complete state characterization using macroscopic analyser regions and subsequent reconstruction \cite{An2024}; metasurface-assisted QKD has generated and distributed hybrid photonic modes while the final protocol measurement remains a separate operation \cite{Jiang2025}. Neither approach makes one subwavelength-interleaved aperture the complete passive-BB84 analyser. A QKD receiver must directly realize the prescribed positive-operator-valued measure (POVM), because its raw outcome probabilities, efficiency mismatch, and crosstalk determine the quantum bit error rate (QBER) and the resulting bound on the secure key rate.

Here, we make the metasurface itself the measurement device and evaluate it as a direct four-outcome BB84 receiver. We analyze wavelength-specific silicon designs at 780, 1550, 2000, and 10\,600~nm, together with additional complete candidates at 780 and 1550~nm. Each aperture interleaves phase functions for linear and circular polarization so that horizontal ($H$), vertical ($V$), right-circular ($R$), and left-circular ($L$) outcomes occupy four focal-plane regions. We obtain conditional response matrices from the simulated two-dimensional focal-plane intensities, compare them with the ideal passive BB84 response, and propagate the matched-basis errors and simulated throughput into a device-level key model.  
Detector regions are determined independently at each wavelength using the same quantitative integration-region selection method, thereby preventing arbitrary post-processing choices from biasing comparisons between designs.

We trace the receiver operation from nanobar geometry to quantum-measurement performance. At each wavelength, a nanobar lookup sweep defines a discrete phase library, from which a subwavelength-interleaved aperture is assembled to direct the four BB84 (or indeed BBM92) outcomes. The resulting full-wave focal fields are converted into conditional detection probabilities and, subsequently, into protocol-level performance metrics. This process is applied independently at 780, 1550, 2000, and 10600 nm, spanning the near-infrared, telecommunications, short-wave-infrared, and mid-infrared regimes. We emphasize that these are distinct wavelength-specific designs rather than a single achromatic metasurface.

\section{Results and Discussion}

\subsection{Protocol measurement and detector-port integration}

The incident polarization qubit is written in the basis $\{\ket{H},\ket{V}\}$. We use $\ket{R}=(\ket{H}+i\ket{V})/\sqrt{2}$ and $\ket{L}=(\ket{H}-i\ket{V})/\sqrt{2}$; these definitions fix the circular-polarization convention used throughout this work. Equal-probability passive basis selection is represented by the POVM elements
\begin{equation}
\begin{aligned}
\hat{E}_H&=\tfrac12\proj{H},&\quad \hat{E}_V&=\tfrac12\proj{V},\\
\hat{E}_R&=\tfrac12\proj{R},& \hat{E}_L&=\tfrac12\proj{L},
\end{aligned}
\label{eq:povm}
\end{equation}
for which $\sum_m \hat{E}_m=\identity$, the identity matrix. We note that the factor $1/2$ is the basis-selection weight, which does not introduce any additional insertion loss. Consequently, with input and outcome order $H,V,R,L$, the ideal conditional response matrix is
\begin{equation}
C^{\rm ideal}=\begin{pmatrix}
1/2&0&1/4&1/4\\
0&1/2&1/4&1/4\\
1/4&1/4&1/2&0\\
1/4&1/4&0&1/2
\end{pmatrix}.
\label{eq:idealresponse}
\end{equation}
Here, \[
C_{i,m}=P(m\mid i),
\]where \(i\) is the polarization state sent into the metasurface and \(m\) is the detector outcome \(H,V,R,\) or \(L\).
Incompatible-basis events are therefore expected to be random.  

\begin{figure*}[!tbp]
\centering
\includegraphics[width=0.698\textwidth]{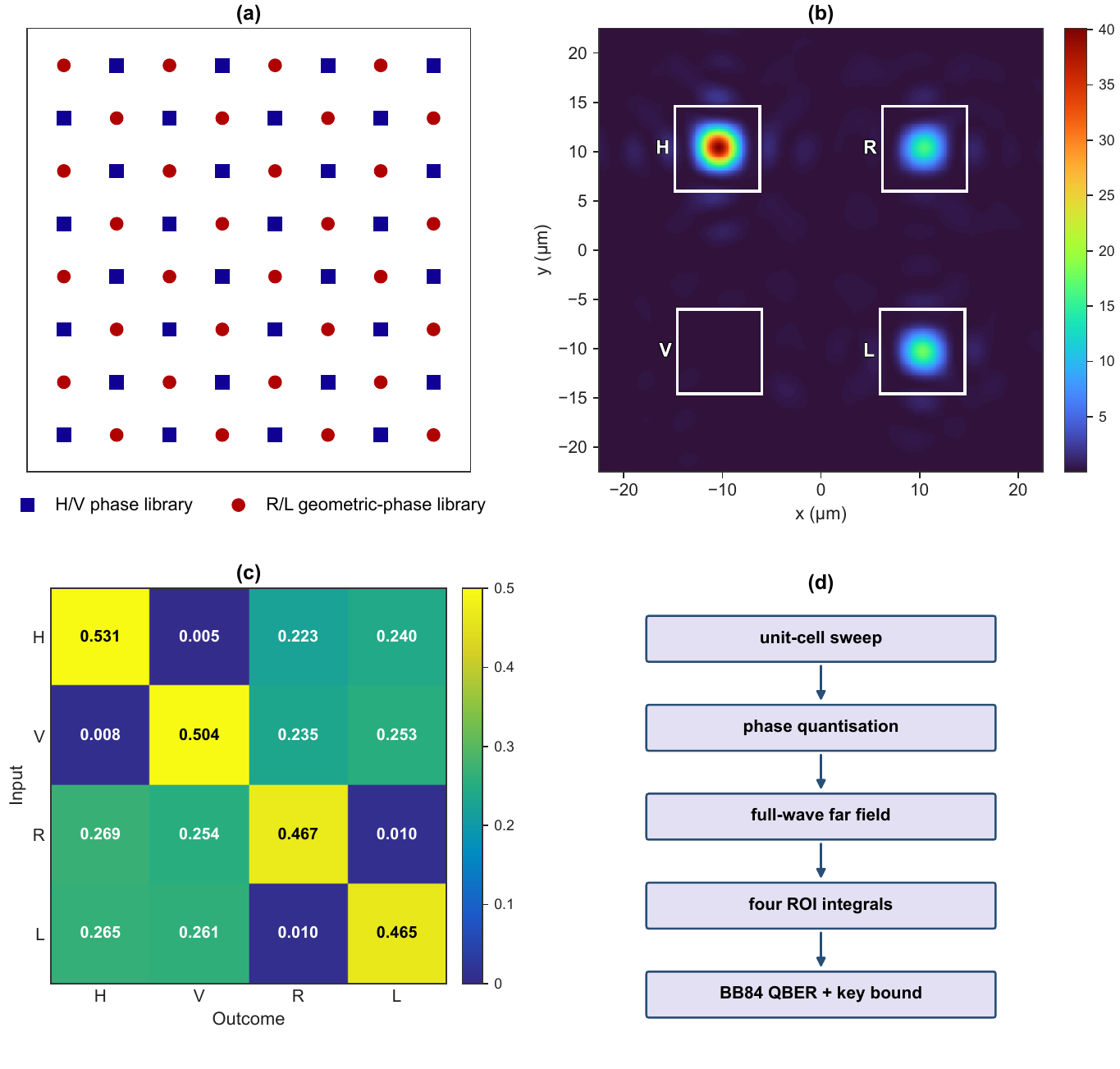}
\caption{\textbf{Focal-plane reconstruction of the BB84 measurement.} \textbf{(a)} Interleaved linear- and circular-basis sublattices. \textbf{(b)} Selected 1550~nm $H$-input far field with the calibration-fixed four-region integration boundaries. \textbf{(c)} Conditional response obtained from the integrated detector powers, with a probability fidelity of 99.08\%. \textbf{(d)} Analysis sequence from the unit-cell sweep to the BB84 metrics. Each port centre is obtained from a local elliptical-Gaussian-plus-constant fit to its matched-state spot. One common square per device is the smallest that retains at least 99\% of every fitted core and 95\% of the empirical background-corrected energy, with no more than 10\% fitted-baseline contribution. Raw intensity is then integrated, the four powers are row-normalized, and physical spots are assigned to $H,V,R,L$ by fitting the complete four-input response to the ideal passive BB84 matrix.}
\label{fig:method}
\end{figure*}

For each simulated focal-plane intensity $I_i(x,y)$, four region integrals are computed,
\begin{equation}
S_{i,k}=\sum_{(r,c)\in\Omega_k} I_i(r,c),\qquad
C_{i,m}=\frac{S_{i,\pi(m)}}{\sum_{k=1}^{4}S_{i,k}},
\label{eq:integration}
\end{equation}
where $i$ labels the input state, $\Omega_k$ is a fixed detector region after calibration, and $\pi$ maps physical regions to logical outcomes. Four $41\times41$-pixel regions centred near indices 50 and 150 are used only as broad initialization windows and as a reference comparison. Within each window, an axis-aligned elliptical Gaussian plus constant baseline is robustly fitted to the matched-state intensity. 
A single integer half-width is imposed on all four ports of a device to avoid outcome-dependent apertures. It is chosen as the smallest size satisfying three numerical criteria for every port: at least 99\% fitted-core retention, at least 95\% of the background-corrected energy in a generous 71-pixel reference square, and at most 10\% estimated constant-baseline power. The resulting square sides are 23, 39, 29, and 21 pixels at 780, 1550, 2000, and 10\,600~nm, respectively; their largest fitted-baseline fraction is 4.9\%. All 24 port permutations are then evaluated, and the single mapping minimizing the response-matrix RMSE is retained. We note that this mapping determines integration area labels only, and neither fitted model nor background subtraction enters the final raw-power sums in the far-field analysis.

The four-region probability fidelity for state $i$ is the squared Bhattacharyya coefficient,
\begin{equation}
F_i=\left(\sum_m\sqrt{C_{i,m}C^{\rm ideal}_{i,m}}\right)^2,
\quad \overline{F}=\tfrac14\sum_iF_i.
\label{eq:fidelity}
\end{equation}
Absolute throughput is reported separately. If $|T_i|$ is the simulated transmission magnitude and $f_i^{\rm ROI}$ is the fraction of the full sampled focal plane contained in the four regions, then $\eta_i^{\rm coll}=|T_i|f_i^{\rm ROI}$. This allows us to evaluate transmission and collection efficiency separately, which is useful because normalization removes the total detected power. 
A useful receiver must produce the correct relative detection probabilities while also delivering sufficient optical power to the detectors. 

\subsection{Design process: Nanobar sweeps to interleaved four-focus aperture}

The design workflow begins with periodic-boundary unit-cell simulations. Rectangular dielectric nanobars are swept in length and width to tabulate transmission amplitude and phase for orthogonal linear inputs. The phase maps are folded into $0$--$2\pi$. For the linear-polarization sublattice, pairs of dimensions are selected to approximate independently specified $x$- and $y$-phase values. For the circular-polarization sublattice, nanobars are first filtered for approximately $\pi$ retardance; their propagation phase and in-plane rotation then supply the mean and differential phases required for opposite circular inputs. Eight discrete phase levels are used in each phase library.

Four continuous focusing profiles are calculated from
\begin{equation}
\phi_m(x,y)=-\frac{2\pi}{\lambda}
\left[\sqrt{(x-x_m)^2+(y-y_m)^2+f^2}-f\right],
\label{eq:focusing}
\end{equation}
and quantized to the available libraries. The linear and circular cells occupy complementary parity sites of the same checkerboard aperture. For the selected devices, the periods 380, 750, 900, and 4700~nm correspond to $p/\lambda=0.487$, 0.484, 0.450, and 0.443, respectively. The basis identity therefore alternates at less than half a wavelength, while the repeat distance $\sqrt{2}p$ of either parity sublattice remains below one wavelength. This avoids an architecture which is essentially four macroscopic polarimeters placed side by side, and the illuminated aperture jointly contributes to four protocol outcomes, offering no propagating-wave-resolvable region belonging to only one basis. This procedure is applied independently to each wavelength-specific design. Here and throughout, the label P\(p\) identifies the unit-cell period \(p\) in nanometres; for example, P380 denotes a design with \(p=380~\mathrm{nm}\). For each design, the phase-and-transmission library was generated by varying the nanobar length \(L\) and width \(W\) while holding the wavelength, material, unit-cell period and nanobar height fixed.

\begin{table*}[!t]
\centering
\caption{Selected wavelength-specific design parameters.}
\label{tab:designs}
\small
\begin{tabular}{@{}lcccccc@{}}
\toprule
$\lambda$ & Material & Period (nm) & Height (nm) & Radius ($\mu$m) & $f$ ($\mu$m) & Spot offsets ($\mu$m) \\
\midrule
780 nm & Si & 380 & 450 & 9.31 & 20 & $\pm4.5$ \\
1550 nm & Si & 750 & 1000 & 22.125 & 100 & $\pm10$ \\
2000 nm & Si & 900 & 1100 & 28.8 & 100 & $\pm12.5$ \\
10.6 $\mu$m & Si & 4700 & 10\,000 & 117.5 & 150 & $\pm60$ \\
\bottomrule
\end{tabular}
\end{table*}

\Cref{tab:designs} summarizes the selected geometries. At 780~nm, the Si P380 library sweeps length and width from 20 to 370~nm on a $36\times36$ grid, while the TiO$_2$ P360 candidate uses 10--350~nm on a $35\times35$ grid. At 1550~nm, the P700 and P750 candidates use 100--650~nm on a $56\times56$ grid and 200--700~nm on a $51\times51$ grid, respectively. The 2~$\mu$m P900 design uses a 200--800~nm, $61\times61$ library, and the current 10.6~$\mu$m P4700 design uses a 1--4~$\mu$m, $101\times101$ library. The full candidate comparison is provided in the Supporting Information.

\subsection{Multiwavelength spatial separation of the two BB84 bases}

\begin{figure*}[!tbp]
\centering
\includegraphics[width=0.96\textwidth]{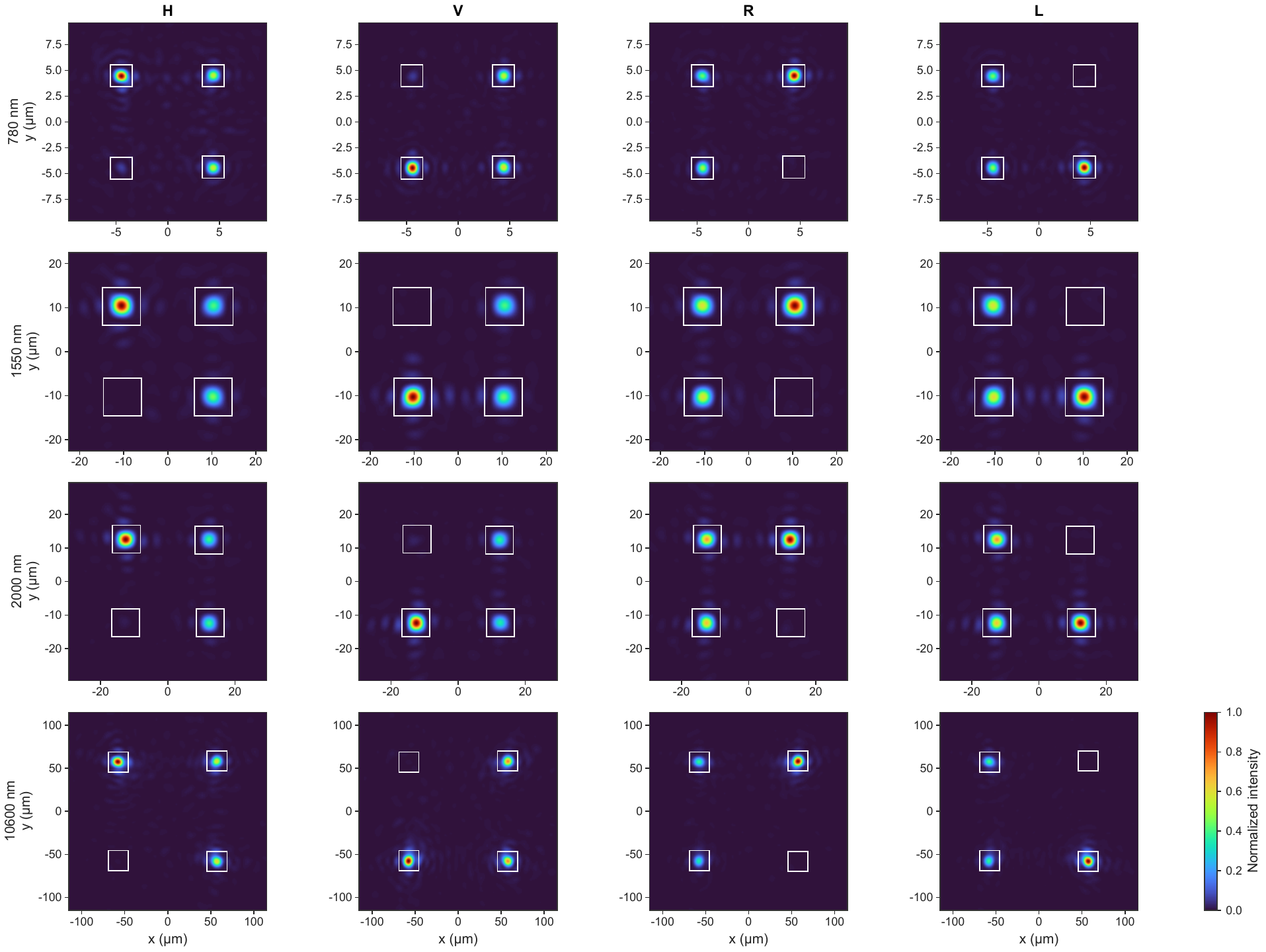}
\caption{\textbf{Full-wave focal-plane intensity for the selected four-port designs.} Rows show independent metasurfaces optimized at 780, 1550, 2000, and 10\,600~nm; columns show $H$, $V$, $R$, and $L$ inputs. Intensities are normalized within each panel for spatial comparison, while absolute transmission and collection are retained separately in the quantitative analysis. Rectangles show the calibration-fixed detector regions used in \Cref{eq:integration}.}
\label{fig:farfields}
\end{figure*}

The full-wave focal planes in \Cref{fig:farfields} show that all four wavelength-specific apertures generate the intended quadrant-dependent response. For a matched-basis input, one of its two logical ports dominates and its orthogonal port is suppressed. Power is nevertheless divided between the incompatible basis ports, as required by \Cref{eq:idealresponse}. This provides a more stringent reading of the fields than selecting the visually brightest spot: a useful design must reproduce all sixteen input--outcome probabilities.

The spot widths and background structure differ between designs. The 780~nm case has the largest leakage into its orthogonal matched-basis port. The 1550~nm field is the cleanest overall, whereas the 2~$\mu$m and 10.6~$\mu$m fields retain low matched-basis leakage but exhibit more unequal incompatible-basis splitting. The full response matrices quantify these features across all four measurement outcomes.

\subsection{Simulated response matrices and intrinsic QBER}

\begin{figure*}[!tbp]
\centering
\includegraphics[width=0.8\textwidth]{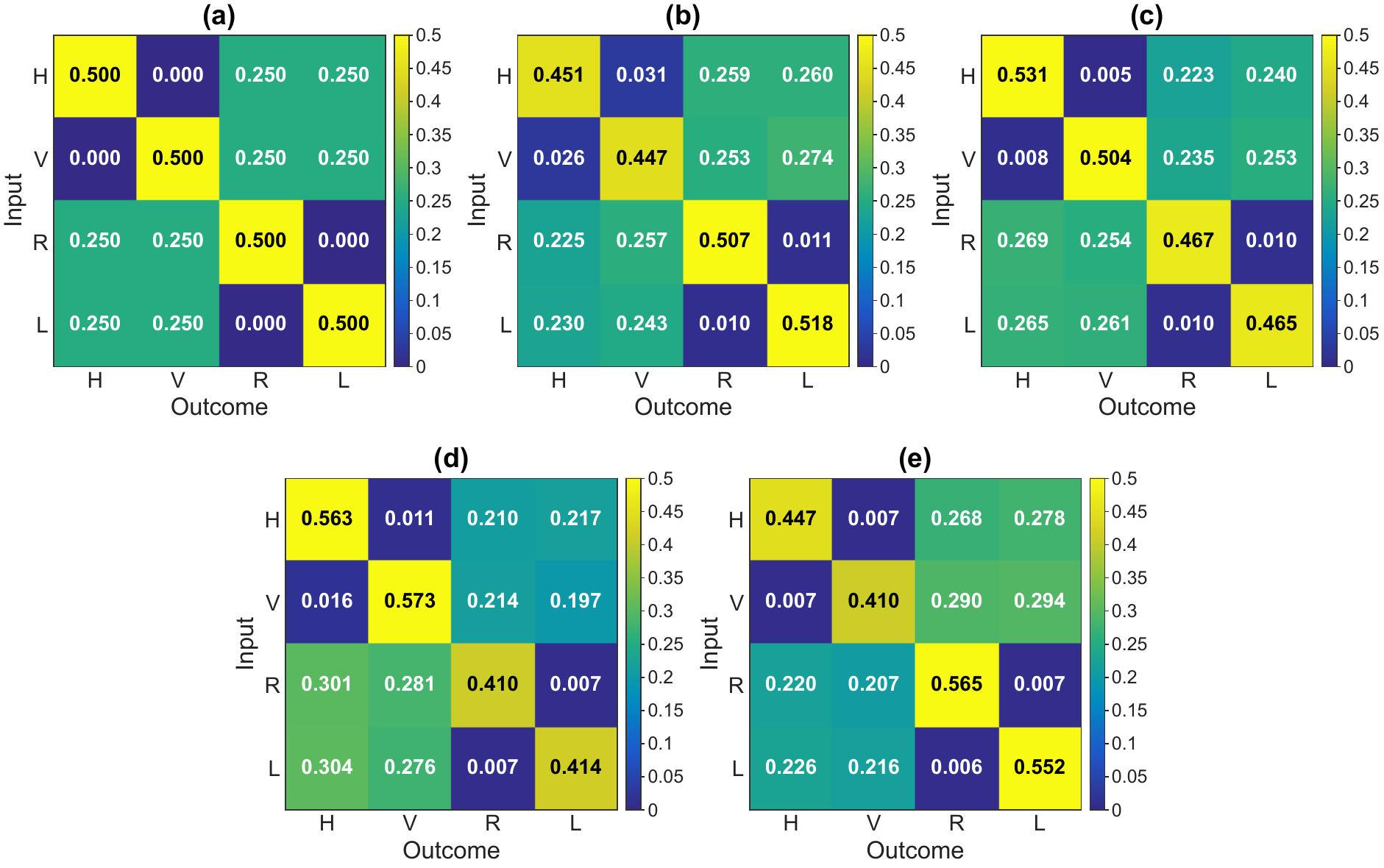}
\caption{\textbf{Conditional four-outcome response matrices.} \textbf{(a)} Ideal passive BB84 response. \textbf{(b--e)} Responses calculated from the selected full-wave intensities at 780, 1550, 2000, and 10\,600~nm, respectively, with probability fidelities of 0.980, 0.991, 0.983, and 0.989. Rows are input states and columns are detector outcomes in the order $H,V,R,L$; every row sums to unity over the four accepted regions. Diagonal entries are matched outcomes, the orthogonal port in the same basis represents a bit error, and the two outcomes in the conjugate basis form the ideally balanced random split.}
\label{fig:matrices}
\end{figure*}

The simulated matrices in \Cref{fig:matrices} closely follow the non-identity target. At 1550~nm, for example, the $H$ and $V$ matched-port probabilities are 0.531 and 0.504, while the orthogonal leakages are 0.005 and 0.008. The $R$ and $L$ matched-port probabilities are 0.467 and 0.465 with orthogonal leakages of 0.010 in each case. The residual deviations occur mainly as imbalance among the two incompatible-basis outcomes and as weak matched-basis leakage.

Let $N_{HV}=\sum_{i,j\in\{H,V\}}C_{i,j}$ and $N_{RL}=\sum_{i,j\in\{R,L\}}C_{i,j}$ denote the accepted probability within each basis. The device-induced errors conditioned on a retained basis are
\begin{align}
Q_{HV}&=\frac{C_{H,V}+C_{V,H}}{N_{HV}},\nonumber\\
Q_{RL}&=\frac{C_{R,L}+C_{L,R}}{N_{RL}}.
\label{eq:qber}
\end{align}
The mean intrinsic QBER is $(Q_{HV}+Q_{RL})/2$. It is 3.96\% at 780~nm, 1.68\% at 1550~nm, 2.01\% at 2~$\mu$m, and 1.42\% at 10.6~$\mu$m. The individual basis errors, fidelities, and throughputs are given in \Cref{tab:metrics}.

\begin{figure*}[!tbp]
\centering
\includegraphics[width=0.8\textwidth]{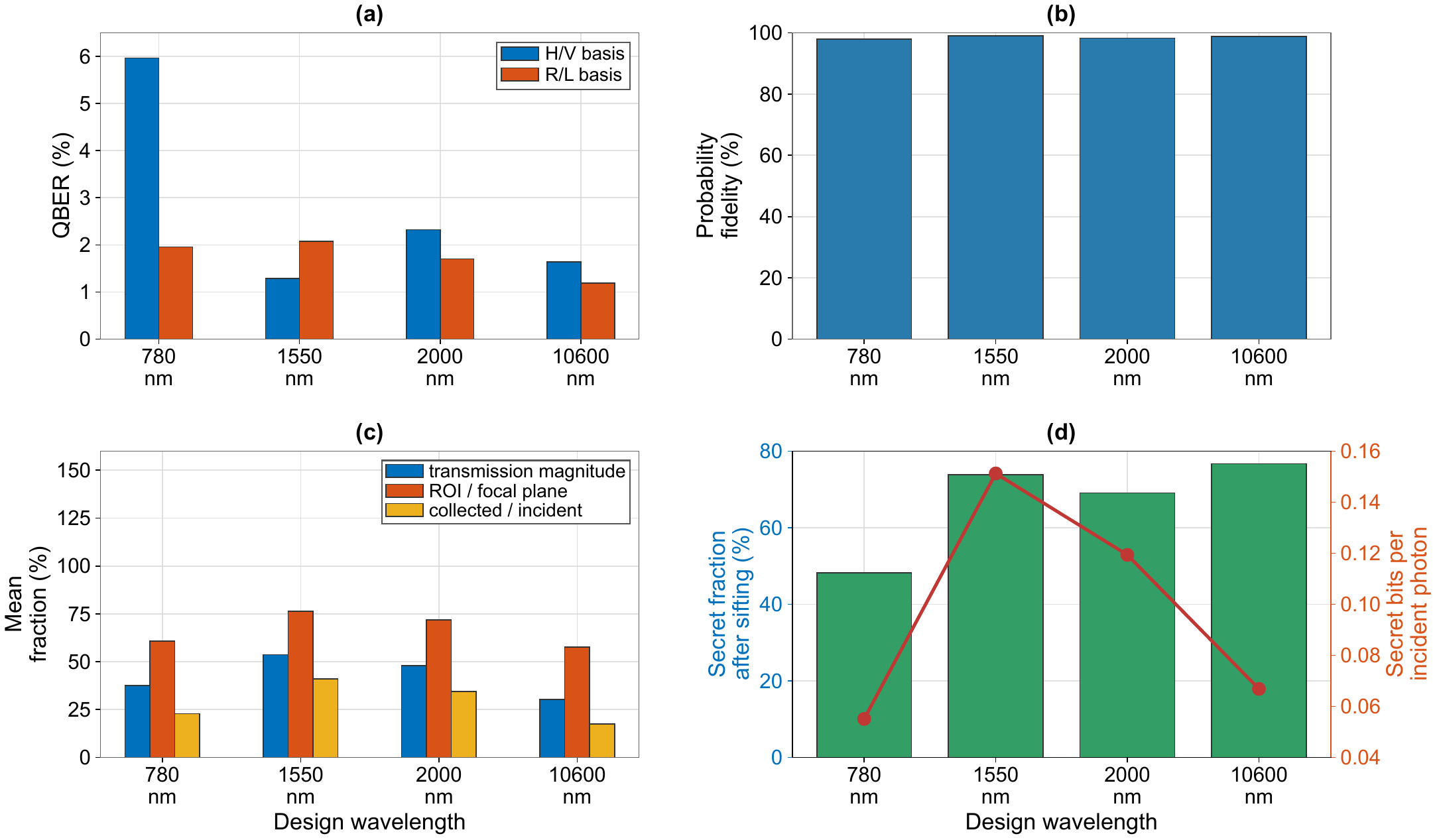}
\caption{\textbf{Optical and protocol metrics for the selected designs.} \textbf{(a)} Basis-conditioned QBER. \textbf{(b)} Conditional probability fidelity. \textbf{(c)} Mean transmission, four-region focal-plane collection fraction, and collected efficiency. \textbf{(d)} Secret fraction after sifting and secret bits per incident photon under the asymptotic model in \Cref{eq:key}.}
\label{fig:metrics}
\end{figure*}

\begin{table*}[!t]
\centering
\caption{Performance of the selected designs. Efficiencies are means over $H,V,R,L$. The secret fraction and yield are device-level asymptotic values with error-correction inefficiency $f_{\rm EC}=1.16$ (assuming practical error correction publicly reveals 16\% more information than the minimum required) \cite{Ma2005RealisticDevices} and exclude channel, source, detector, and finite-key effects.}
\label{tab:metrics}
\small
\begin{tabular}{@{}lcccccccc@{}}
\toprule
$\lambda$ & $Q_{HV}$ & $Q_{RL}$ & $\overline F$ & $|T|$ & $f^{\rm ROI}$ & $\eta^{\rm coll}$ & $r_{\rm sec}$ & $Y_{\rm sec}$ \\
\midrule
780 nm & 5.97\% & 1.96\% & 97.96\% & 37.55\% & 60.78\% & 22.82\% & 48.26\% & 0.0551 \\
1550 nm & 1.29\% & 2.08\% & 99.08\% & 53.63\% & 76.34\% & 40.94\% & 73.88\% & 0.1513 \\
2000 nm & 2.32\% & 1.70\% & 98.30\% & 48.01\% & 71.94\% & 34.54\% & 69.09\% & 0.1193 \\
10.6 $\mu$m & 1.64\% & 1.19\% & 98.85\% & 30.20\% & 57.71\% & 17.43\% & 76.68\% & 0.0668 \\
\bottomrule
\end{tabular}
\end{table*}

The strongest conditional fidelity is obtained by the 1550~nm P750 design (99.08\%) with a mean QBER of 1.68\%. P700 collects more power (46.53\% versus 40.94\%) but has 2.52\% mean QBER under its larger calibration-defined square; under the optical benchmark of \Cref{eq:key}, P750 gives the higher device-level secret yield (0.1513 versus 0.1471 bits per incident photon) and is selected for the four-wavelength comparison. At 780~nm, the silicon P380 design has lower collected efficiency than the TiO$_2$ P360 candidate (22.82\% versus 31.50\%) but substantially better adherence to the target probability distribution; its predicted secret yield is correspondingly higher (0.0551 versus 0.0516). These comparisons demonstrate why the selection objective must combine protocol error and throughput. The reconstructed-operator comparison in \Cref{sec:characterizedsecurity} evaluates all six candidates with their characterized basis-dependent responses.

\subsection{Device-level secret fraction and robustness}

For the equal-basis BB84 protocol considered here, a lossless, error-free receiver gives $Y_{\rm ideal}=1/2$ secret bit per incident photon: half the photons survive basis matching, and each retained photon supplies one secret bit. With polarization-independent collection efficiency $\eta$, the corresponding error-free yield is $\eta/2$. For the simulated devices, we calculate the actual mean matching-basis detection probability $\overline p_{\rm sift}$, which includes transmission loss, finite detector-region collection and the realized basis-selection weights.

By the term \emph{optical benchmark}, we denote the earlier four-state QBER-based estimate using these simulated optical responses.  It applies the following common secret fraction to the sifted detections:
\begin{equation}
r_{\rm sec}=\max\left\{0,1-f_{\rm EC}\binaryentropy(Q_{HV})
-\binaryentropy(Q_{RL})\right\},
\label{eq:key}
\end{equation}
where $f_{\rm EC}=1.16$ is the error-correction inefficiency factor, and $\binaryentropy$ is the binary entropy function \cite{ShorPreskill2000}. Error correction reconciles disagreements between Alice's and Bob's retained bit strings. We adopt the factor 1.16 from the low-error-rate values tabulated by Ma \cite[Table~4.1]{Ma2005RealisticDevices}: practical error correction is assumed to reveal 16\% more information publicly than the minimum required to correct the same errors. The public information disclosed during reconciliation reduces the secret-key length; the term $f_{\rm EC}\binaryentropy(Q_{HV})$ budgets this leakage in the reference expression. Privacy amplification then shortens the reconciled key to remove Eve's bounded information. The term $\binaryentropy(Q_{RL})$ estimates that deduction using the opposite-basis QBER as a phase-error surrogate.

The benchmark yield is $Y_{\rm sec}=\overline p_{\rm sift}r_{\rm sec}$. For P750, the lossless, error-free reference is 0.500, while the actual receiver gives $\overline p_{\rm sift}=0.20472$ matching detections per incident photon after collection losses and basis sifting. The secret fraction $r_{\rm sec}=0.73884$ then gives $Y_{\rm sec}=0.15126$ after the reconciliation and privacy-amplification deductions. Loss accounts for the larger reduction in this example, but both loss and errors enter the benchmark. The metric is evaluated at the device plane with ideal single-photon preparations; channel attenuation, detector inefficiency, dark counts, multiphoton emission, decoy-state estimation and finite-key terms are outside this model.

\begin{figure}[!tbp]
\centering
\includegraphics[width=0.48\textwidth]{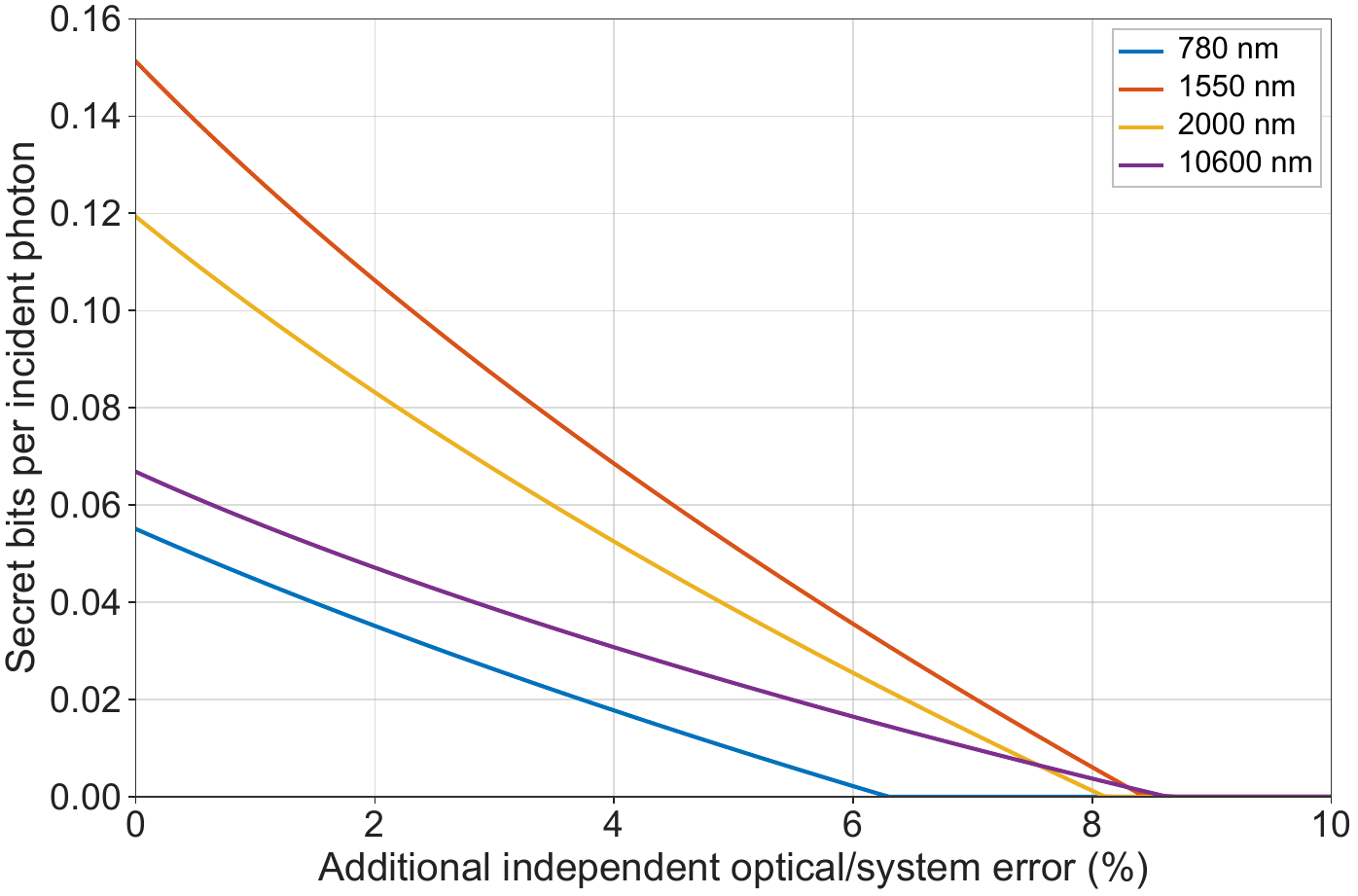}
\caption{\textbf{Effect of additional bit errors on secret-key yield.} The curves show how the number of secret bits per incident photon decreases as additional errors are introduced into each device's simulated measurement outcomes. The added errors flip a bit from 0 to 1 or from 1 to 0 with equal probability. }
\label{fig:robustness}
\end{figure}

The resulting yields are 0.0551, 0.1513, 0.1193, and 0.0668 secret bits per incident photon at 780, 1550, 2000, and 10\,600~nm, respectively. \Cref{fig:robustness} applies an additional symmetric bit-flip channel to the simulated responses. The 1550~nm, 2~$\mu$m, and 10.6~$\mu$m designs have the largest error budgets, whereas the 780~nm design loses positive yield first because its nominal $H/V$ leakage is highest.

To test how sensitive the results are to the detector-region size, we expanded and contracted each calibrated square by moving its boundaries outwards and inwards by 2 and 5 pixels, keeping its fitted centre fixed. We find that enlarging the regions increases collected power but also admits more halo and background, so QBER generally rises. At 1550~nm, the tested range moves the mean QBER from 0.96\% to 2.57\% while the key yield remains between 0.140 and 0.165. At 780~nm, it spans 3.00--5.26\% mean QBER and 0.0456--0.0595 yield. Detector aperture is therefore considered an optical/security co-design parameter. 

\subsection{Security with reconstructed measurement operators}
\label{sec:characterizedsecurity}

The additional $\ket{D}=(\ket{H}+\ket{V})/\sqrt2$ and $\ket{A}=(\ket{H}-\ket{V})/\sqrt2$ simulations complete six-state detector tomography. All inputs are integrated over the same four physical detector regions. Their absolute probabilities per incident photon determine the reconstructed operators $\hat E_j^{\rm rec}$, including a no-click operator $\hat E_{\loss}^{\rm rec}=\identity-\sum_j\hat E_j^{\rm rec}$. We constrain the reconstructed operators to be positive and account for the simulated polarization-dependent loss, unequal basis weights and wrong-port leakage. We then compare them with the ideal operators in \Cref{eq:povm}.

\begin{figure*}[!tbp]
\centering
\includegraphics[width=0.65\textwidth]{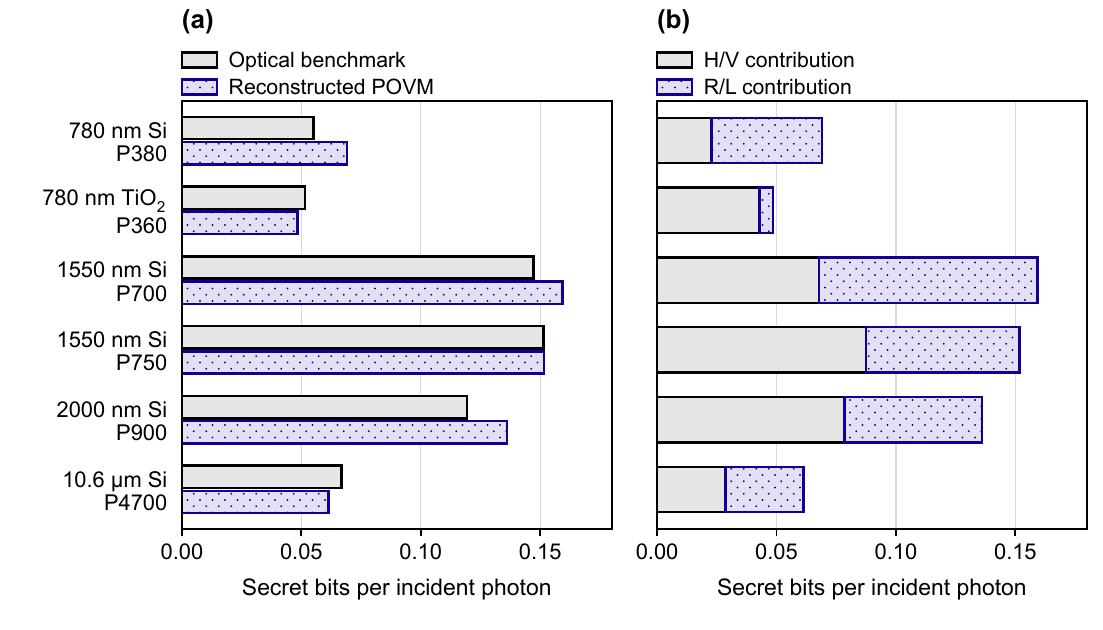}
\caption{\textbf{Secret-key yields with characterized measurement operators.} \textbf{(a)} The \emph{optical benchmark} estimates the secret bits per incident photon from the four H, V, R and L input-state simulations using \Cref{eq:key}. It accounts for optical loss and the key-length reductions needed to reconcile bit errors and remove Eve's bounded information through privacy amplification. The \emph{reconstructed POVM} bars give the lower bound from \Cref{eq:tomographickey}, using the receiver's measurement operators fitted to all six input states and a separate privacy bound for each key basis. This calculation can give a higher or lower yield than the benchmark. Both use the original physical detector regions. \textbf{(b)} The stacked segments show the secret bits per incident photon contributed by matching detections in the linear (H/V) and circular (R/L) bases. Each contribution is averaged over Alice's equally likely basis choices and includes the probability of a matching-basis detection and the fraction of detected bits remaining secret after error correction and privacy amplification. Adding the two contributions gives the reconstructed-POVM total in (a). Values are asymptotic yields for ideal single-photon preparations and a trusted receiver at its nominal operating probabilities, with error-correction inefficiency $f_{\rm EC}=1.16$. The lossless, error-free yield for this protocol is 0.5, outside the plotted range. Reconstruction residuals and specified-operator-uncertainty bounds are given in the Supporting Information.}
\label{fig:characterizedsecurity}
\end{figure*}

For a trusted polarization-qubit receiver with ideal single-photon preparations, the security calculation treats H/V and R/L as separate, publicly identified key bases. Let $b=Z$ denote H/V and $b=Y$ denote R/L. The matching-click probability $p_b$ is conditioned on Alice's choice of basis, and $e_b$ is its bit-error probability. 

We use a virtual filter, a mathematical description of which detection events are retained, to account for the receiver's state-dependent acceptance. Combining the reconstructed measurement operators with the correlations in the complementary basis gives a bound on the phase-error probability $e_{{\rm ph},b}$. This describes the errors that would occur in a hypothetical complementary-basis test of the retained signals.

The uncertainty relation with quantum memory then bounds how uncertain Eve is about Alice's retained bits, even when Eve holds quantum systems that she can measure later \cite{Berta2010}:
\begin{equation}
H(A_b\mid E,{\rm pass}_b)\geq 1-h_2(e_{{\rm ph},b}).
\end{equation}
Here, $b$ denotes the key basis, $A_b$ is Alice's key bit in that basis, and $E$ represents Eve's quantum information. The condition ${\rm pass}_b$ means that the receiver recorded an accepted matching-basis detection. The conditional von Neumann entropy $H(A_b\mid E,{\rm pass}_b)$ measures Eve's remaining uncertainty about Alice's bit, in bits per accepted event, and $h_2$ is the binary entropy function. The inequality therefore gives a lower bound on that uncertainty before information is disclosed during error correction.

\begin{align}
r_b&=\bigl[1-h_2(e_{{\rm ph},b})-1.16h_2(e_b)\bigr]_+,\\
Y_{\rm tom}&=\tfrac12p_Zr_Z+\tfrac12p_Yr_Y.
\label{eq:tomographickey}
\end{align}
Here $[x]_+=\max(0,x)$, and the factors $1/2$ are Alice's equally likely basis choices; the receiver's basis-selection probabilities are already contained in $p_b$. Thus both matched bases contribute to the yield, with no assumption of an exactly 50:50 receiver or identical error rates. In each basis, $1.16h_2(e_b)$ accounts for reconciliation leakage, while $h_2(e_{{\rm ph},b})$ determines the privacy-amplification deduction from the operator-derived complementary bound.
The reconstructed-POVM yield is to be understood as an asymptotic lower bound for the characterized receiver at its nominal operating probabilities. 

We first estimate the secret-key yield from the H, V, R and L simulations, using the simulated collection losses and quantum bit error rates. This four-state QBER-based estimate applies the BB84 secret-fraction expression below to the matching-basis detections.
 Thus, this optical benchmark applies one common QBER-based secret fraction, whereas \Cref{eq:tomographickey} uses separate bit errors, operator-derived phase errors and acceptance weights for H/V and R/L. It also uses the fitted six-state probabilities. Consequently, the reconstructed-POVM result can lie above or below the benchmark. 
 The complete derivation, operator coefficients and six-state probabilities are given in the Supporting Information.

All six reconstructed receivers give positive yields (\Cref{fig:characterizedsecurity}). For the four designs emphasized above, $Y_{\rm tom}$ is 0.06901, 0.15164, 0.13599 and 0.06130 bits per incident photon at 780, 1550, 2000 and 10\,600~nm, respectively. The P750 result closely matches its original 0.15126 optical benchmark. The additional 1550~nm P700 and 780~nm TiO$_2$ P360 candidates give 0.15935 and 0.04846, respectively; P700 therefore has the highest yield under the characterized-receiver bound. These positive bounds include the imperfections represented by the reconstructed POVMs. They apply to the fitted nominal receiver models under asymptotic collective attacks with authenticated public communication; the Supporting Information reports raw-data consistency and operator-uncertainty sensitivity separately.

\subsection{Subwavelength basis co-location and receiver integrity}

The central functional result of our work is a direct mapping between a phase-engineered dielectric aperture and a QKD measurement. The output is already a four-symbol protocol outcome, and no Stokes-vector inversion is required to assign a click. This distinguishes the device from a general polarimeter and from a metasurface used only upstream for state preparation or mode multiplexing. The multiwavelength set also shows that the workflow is scalable across materially different operating regimes: the unit-cell library and focal geometry change, but the target measurement and validation remain the same.

The checkerboard interleaving adds a receiver-integrity property that is absent when measurement bases occupy separately addressable macroscopic regions. In a spatially segmented analyser, a basis-selective mask can in principle expose, illuminate, attenuate, or block one analyser region independently of the other. Here, the linear- and circular-basis functions alternate at $p<\lambda/2$ across the full aperture. A conventional far-field mask cannot resolve and transmit only one parity sublattice; a perforated mask patterned at the cell scale would require subwavelength openings and would strongly attenuate or scatter the incident field rather than provide an independent propagating channel to one basis. The architecture therefore removes the straightforward spatial-filtering route by which an adversary could address the basis analysers separately. In this precise sense, subwavelength basis co-location is, in addtion to being a compactness feature, a security-relevant physical constraint on spatial access.

The six-state characterization in \Cref{sec:characterizedsecurity} resolves the polarization-qubit measurement at the nominal wavelength and admitted spatial mode of each device. The trusted-receiver model includes the reconstructed efficiency mismatch; spectral, fabrication and incident-mode variations are distinct from this fixed-mode characterization \cite{Makarov2006,Beaudry2008}. Subwavelength basis co-location supplies the spatial-access constraint described above, while the key bound follows from the characterized operators and complementary correlations.

Although the present study does not model an entangled-photon source or paired receivers, the same local measurement can in principle be used for BBM92 by placing one receiver in each arm of an entangled pair. If $\hat{E}_m^A$ and $\hat{E}_n^B$ are the experimentally bounded local effects, the joint statistics are $p_{mn}=\Tr[(\hat{E}_m^A\otimes \hat{E}_n^B)\hat{\rho}_{AB}]$ \cite{BennettBrassardMermin1992}.   

\FloatBarrier
\section{Conclusion}

We have designed and evaluated four wavelength-specific dielectric metasurface receivers that directly implement the $H/V$ and $R/L$ measurement bases of passive BB84 within one subwavelength-interleaved aperture. The complete process links the unit-cell transmission/phase sweep, discrete linear- and circular-polarization libraries, interleaved four-focus layout, full-wave propagation, calibrated-region integration, and logical port assignment directly to QKD performance metrics.

Across 780~nm to 10.6~$\mu$m, the selected designs achieve 97.96--99.08\% conditional-probability fidelity and 1.42--3.96\% mean intrinsic QBER. Under the original optical benchmark, the 1550~nm P750 design provides the best throughput--error compromise, combining 40.94\% four-region collected efficiency with a device-level asymptotic yield of 0.151 secret bits per incident photon. Six-state reconstruction gives 0.15164 for P750 and 0.15935 for P700 under the characterized-receiver bound. The analysis also shows that probability fidelity alone is insufficient, since candidate ranking changes when absolute throughput and a quantitatively defined detector aperture are propagated into the secure-key-rate key metric.

The six-state analysis gives positive asymptotic secret-key bounds for all six fitted receiver models while retaining their polarization-dependent loss, basis imbalance and crosstalk. Both H/V and R/L matched outcomes contribute to the key. 

\section{Methods}

\subsection{Numerical datasets and candidate selection}

Complete full-wave datasets were analyzed for 780~nm (Si P380 and TiO$_2$ P360), 1550~nm (Si P700 and P750), 2000~nm (Si P900), and 10.6~$\mu$m (Si P4700). Each dataset comprises simulated focal-plane intensity distributions for $H$, $V$, $R$, $L$, $D$, and $A$ input states, together with the transmitted power and spatial coordinate grids. The original four-wavelength comparison selected the largest asymptotic device-level yield under \Cref{eq:key}; conditional probability fidelity served as the secondary criterion when the predicted yields were zero. The additional six-state security analysis retains and compares all six candidates.

\subsection{Electromagnetic workflow}

Full-wave electromagnetic simulations were performed in Ansys Lumerical FDTD for the nanobar lookup and complete-aperture propagation. Nanobar length and width were swept under periodic in-plane boundaries to obtain transmission and phase. The selected phase states were substituted into the four focusing profiles of \Cref{eq:focusing}. Complete devices were propagated to their stated focal planes with \texttt{farfieldexact3d}; each sampled plane contains $200\times200$ values of total electric-field intensity together with the spatial coordinates and transmitted power.

\subsection{Image integration and response metrics}

Post-processing and detector-region integration were performed in MATLAB R2023b using a custom analysis routine. The four inclusive pixel regions $[30{:}70,130{:}170]$, $[30{:}70,30{:}70]$, $[130{:}170,130{:}170]$, and $[130{:}170,30{:}70]$ were used as initialization windows and as a reference comparison. Final centres and common square sides were determined by the fit-and-growth-curve method described above. The simulated raw intensity within the calibrated regions was then integrated without model substitution or background subtraction. Conditional probability fidelity was calculated using \Cref{eq:fidelity}; total-variation distance and permutation-fit RMSE were retained as complementary response metrics.

\subsection{QKD calculation}

Basis-resolved device QBERs were calculated with \Cref{eq:qber}. The mean sifted detection probability averages the absolute matching-basis probabilities over the four equally likely input states. Each probability contains $|T_i|$, the four-region fraction of the sampled plane and the realized matching-basis weight; no further passive-basis factor is applied. The optical benchmark secret fraction follows \Cref{eq:key}. Robustness curves apply an independent symmetric flip probability to the two outcomes of each basis and recompute both QBERs and the key expression. No stochastic fabrication model is inferred from this artificial channel. The additional six-state reconstruction and both-basis security calculation use \Cref{eq:tomographickey}; the Python implementation preserves the original physical detector regions, with native-pixel area weights where the D/A coordinate grid differs. All fitted click and no-click operators are constrained to be positive.

\section*{Supporting Information}

The Supporting Information contains the design and sweep parameters, polarization conventions, detector-region calibration, complete response matrices, candidate comparison, detector-aperture sensitivity, D/A focal-plane data, six-state POVM reconstruction, and the security derivation and both-basis key bounds.

\section*{Acknowledgements}

This work was supported by UK Research and Innovation (UKRI) through the Engineering and Physical Sciences Research Council (EPSRC) Doctoral Training Partnership at the University of Glasgow [grant ref: EP/W524359/1] and by the Royal Society (Grant No. RG\textbackslash{}R1\textbackslash{}251474).

We thank Natale Pruiti and Marc Sorel for their support with Lumerical simulations, and Shuhao Wu, Miles Padgett for helpful discussions.

\section*{Data Availability Statement}

The data and analysis code supporting the findings of this study are available from the corresponding author upon reasonable request.

\makeatletter
\let\auto@bib\@empty
\let\auto@bib@innerbib\@empty
\let\thebibliography\NAT@thebibliography
\let\endthebibliography\endNAT@thebibliography
\def\@bibsetup#1{\setlength{\topsep}{0pt}\NATx@bibsetnum{#1}}
\makeatother

\clearpage
\setcounter{section}{0}
\setcounter{subsection}{0}
\setcounter{subsubsection}{0}
\setcounter{equation}{0}
\setcounter{figure}{0}
\setcounter{table}{0}
\renewcommand{\thesection}{\arabic{section}}
\renewcommand{\thesubsection}{\thesection.\arabic{subsection}}
\renewcommand{\thesubsubsection}{\thesubsection.\arabic{subsubsection}}
\renewcommand{\theequation}{S\arabic{equation}}
\renewcommand{\thefigure}{S\arabic{figure}}
\renewcommand{\thetable}{S\arabic{table}}
\renewcommand{\theHsection}{SI.\arabic{section}}
\renewcommand{\theHsubsection}{\theHsection.\arabic{subsection}}
\renewcommand{\theHsubsubsection}{\theHsubsection.\arabic{subsubsection}}
\renewcommand{\theHequation}{SI.\arabic{equation}}
\renewcommand{\theHfigure}{SI.\arabic{figure}}
\renewcommand{\theHtable}{SI.\arabic{table}}
\renewcommand{\arraystretch}{1.15}
\pdfbookmark[0]{Supporting Information}{supporting-information}
\begin{center}
{\Large\bfseries Supporting Information\par}
\vspace{0.6\baselineskip}
{\large\bfseries Secure Quantum Metasurfaces for Direct Quantum Key Distribution Measurements\par}
\vspace{0.6\baselineskip}
Yan He and Adetunmise C. Dada\par
\vspace{0.3\baselineskip}
{\small School of Physics and Astronomy, University of Glasgow, Glasgow, G12 8QQ, United Kingdom\par}
\vspace{0.3\baselineskip}
{\small Correspondence: Adetunmise C. Dada (\href{mailto:adetunmise.dada@glasgow.ac.uk}{adetunmise.dada@glasgow.ac.uk})\par}
\end{center}
\vspace{0.5\baselineskip}
\section{Design set and candidate selection}

Six complete wavelength-specific models were analyzed: 780~nm Si P380 and TiO$_2$ P360, 1550~nm Si P700 and P750, 2000~nm Si P900, and 10.6~$\mu$m Si P4700. Each model provides four simulated $200\times200$ focal-plane intensity distributions for $H$, $V$, $R$, and $L$ input states, together with the transmitted power and spatial coordinate grids. At each wavelength, the design with the largest asymptotic device-level secret yield under the four-state optical benchmark was selected; conditional probability fidelity served as the secondary criterion if all predicted yields were zero. The six-state, characterized-receiver calculation in Section~\ref{sec:si_security_v4} evaluates all six candidates, including both 1550~nm designs.

The complete design procedure comprises: (i) a periodic-boundary unit-cell sweep; (ii) inspection of the full-$2\pi$ phase and transmission maps; (iii) discrete phase-library selection; (iv) calculation of the continuous focusing phases; (v) phase-to-geometry assignment; (vi) complete-aperture construction and field verification; (vii) focal-plane propagation; and (viii) calibrated detector-region integration.

\section{Target measurement and polarization conventions}

The four-state optical benchmark uses $H,V,R,L$. The D/A simulations complete the six-state receiver characterization in Section~\ref{sec:si_security_v4}. We use
\begin{equation}
\ket{R}=\frac{\ket H+i\ket V}{\sqrt2},\qquad
\ket{L}=\frac{\ket H-i\ket V}{\sqrt2}.
\end{equation}
These equations fix the circular-polarization convention used throughout. Interchanging $R$ and $L$ relabels the corresponding output ports without changing the two-basis protocol. The two bases are mutually unbiased because $|\langle H|R\rangle|^2=|\langle H|L\rangle|^2=1/2$ and similarly for $V$.

For equal passive basis probability, the target measurement operators are
\begin{equation}
\hat{E}_H=\tfrac12\proj H,\quad \hat{E}_V=\tfrac12\proj V,\quad
\hat{E}_R=\tfrac12\proj R,\quad \hat{E}_L=\tfrac12\proj L,
\end{equation}
and the ideal response for rows=input and columns=outcome is
\begin{equation}
C^{\rm ideal}=\begin{pmatrix}
0.5&0&0.25&0.25\\
0&0.5&0.25&0.25\\
0.25&0.25&0.5&0\\
0.25&0.25&0&0.5
\end{pmatrix}.
\label{eq:sideal}
\end{equation}
The H/V and R/L states probe the Pauli-$Z$ and Pauli-$Y$ directions, respectively. The D/A states probe Pauli-$X$. Together the six simulated input states determine the complete qubit receiver model, including loss, through the reconstruction in Section~\ref{sec:si_security_v4}.

\section{Nanobar sweep, phase selection, and aperture construction}

The design procedure is described below.

\subsection{Unit-cell lookup}

Rectangular dielectric cells are simulated on a two-dimensional length--width grid. The complex transmission coefficient for a normally incident linear input is reduced to a magnitude map $T_x(L,W)$ and wrapped phase map $\phi_x(L,W)/(2\pi)$. The orthogonal response is obtained by the corresponding rotated/transposed geometry, producing $T_y$ and $\phi_y$. The selected sweep grids are listed in \Cref{tab:sweeps}.

\begin{table}[h]
\centering
\caption{Wavelength-specific unit-cell sweep and focusing parameters.}
\label{tab:sweeps}
\small
\begin{tabular}{@{}lcccccc@{}}
\toprule
$\lambda$ & Candidate & Size sweep & Grid & Period & $f$ & Offsets \\
 & & (nm) & & (nm) & ($\mu$m) & ($\mu$m) \\
\midrule
780~nm & Si P380 & 20--370 & $36^2$ & 380 & 20 & $\pm4.5$ \\
780~nm & TiO$_2$ P360 & 10--350 & $35^2$ & 360 & 20 & $\pm4.5$ \\
1550~nm & Si P700 & 100--650 & $56^2$ & 700 & 100 & $\pm10$ \\
1550~nm & Si P750 & 200--700 & $51^2$ & 750 & 100 & $\pm10$ \\
2000~nm & Si P900 & 200--800 & $61^2$ & 900 & 100 & $\pm12.5$ \\
10.6~$\mu$m & Si P4700 & 1000--4000 & $101^2$ & 4700 & 150 & $\pm60$ \\
\bottomrule
\end{tabular}
\end{table}

\subsection{Linear- and circular-polarization libraries}

For the LP sublattice, the continuous target phases for the two orthogonal linear inputs are each rounded to one of eight levels. The two phase indices address a selected geometry pair that approximates both requirements. For the CP sublattice, cells whose x/y phase difference is close to $\pi$ are selected, eight propagation-phase states are retained, and an in-plane rotation is calculated from the required R/L differential phase. The complete aperture interleaves LP sites on one checkerboard parity and CP sites on the complementary parity.

For the selected 780, 1550, 2000, and 10.6~$\mu$m designs, the lattice periods give $p/\lambda=0.487$, 0.484, 0.450, and 0.443. Hence the basis function changes every $p<\lambda/2$, and identical-parity sites repeat after $\sqrt{2}p=0.69$, 0.68, 0.64, and 0.63$\lambda$, respectively. An ordinary propagating-wave spatial mask cannot resolve this checkerboard parity. A mask attempting to pass only one basis would require subwavelength openings at the surface and would therefore attenuate and scatter the incident field rather than provide independent macroscopic access to that basis. This is the physical basis for the spatial-access argument in the main text; the simulated response matrices themselves remain the quantities used in the QBER and key-yield calculation.

For each outcome $m$, the target focusing phase is
\begin{equation}
\phi_m(x,y)=-\frac{2\pi}{\lambda}
\left[\sqrt{(x-x_m)^2+(y-y_m)^2+f^2}-f\right].
\end{equation}
The focal-plane coordinate assignments are R $(+,+)$, L $(+,-)$, H $(-,+)$, and V $(-,-)$. After array-display conventions and integration indices are accounted for, the exhaustive probability fit maps physical ROI 1, 2, 3, and 4 to H, V, R, and L for every complete dataset.

\section{Far-field propagation and detector-region integration}

Complete-aperture fields were propagated with \texttt{farfieldexact3d} onto a $200\times200$ focal plane, yielding the spatial coordinate vectors, total-field intensity, and transmitted power for each input polarization.

The four initialization regions were the inclusive squares
\begin{equation}
\begin{split}
\Omega_1&=[30{:}70,130{:}170],\quad
\Omega_2=[30{:}70,30{:}70],\\
\Omega_3&=[130{:}170,130{:}170],\quad
\Omega_4=[130{:}170,30{:}70].
\end{split}
\end{equation}
Each contains $41\times41$ pixels. These boxes are used as generous spot-search windows and as a reference comparison, rather than as the final detector apertures. A first pass integrates them and exhaustively scores all 24 region-to-label permutations against \Cref{eq:sideal}. The minimum-RMSE assignment defines the logical $H,V,R,L$ ports for the complete candidate.

The centre of each logical port is then calibrated from the field in which that port is the matched outcome. Within its historical search window, an axis-aligned elliptical Gaussian plus a constant baseline is fitted using a soft-$L_1$ residual. The fitted model is
\begin{equation}
I_{\rm fit}(r,c)=B+A\exp\!\left[-\frac{(r-r_0)^2}{2\sigma_r^2}-\frac{(c-c_0)^2}{2\sigma_c^2}\right].
\label{eq:sfit}
\end{equation}
The continuous centre $(r_0,c_0)$ is rounded only when the final pixel bounds are constructed. One common integer half-width $h$ is imposed on all four ports of a candidate, preventing an outcome-dependent aperture from artificially improving the response.

The selected $h$ is the smallest value for which every port satisfies all three requirements:
\begin{enumerate}
\item the fitted Gaussian core fraction within the square is at least 0.99;
\item the empirical, background-corrected signal is at least 0.95 of that obtained in a generous $71\times71$-pixel reference square centred on the same fit; and
\item the estimated constant-baseline contribution to the selected raw integral is no more than 0.10.
\end{enumerate}
The final integration always uses the simulated raw intensity. Neither the fitted Gaussian nor the estimated background is substituted into or subtracted from the reported port powers. The fits therefore calibrate a common detector geometry.

\begin{table}[h]
\centering
\caption{Adaptive detector apertures for the four selected designs. The final column is the largest fitted-baseline estimate among the four ports.}
\label{tab:adaptiveapertures}
\small
\begin{tabular}{@{}lccc@{}}
\toprule
$\lambda$ & Half-width $h$ & Square side & Maximum baseline fraction \\
\midrule
780~nm & 11 pixels & 23 pixels & 4.00\% \\
1550~nm & 19 pixels & 39 pixels & 4.92\% \\
2000~nm & 14 pixels & 29 pixels & 3.73\% \\
10.6~$\mu$m & 10 pixels & 21 pixels & 4.91\% \\
\bottomrule
\end{tabular}
\end{table}

\begin{figure}[!htbp]
\centering
\includegraphics[width=0.84\textwidth]{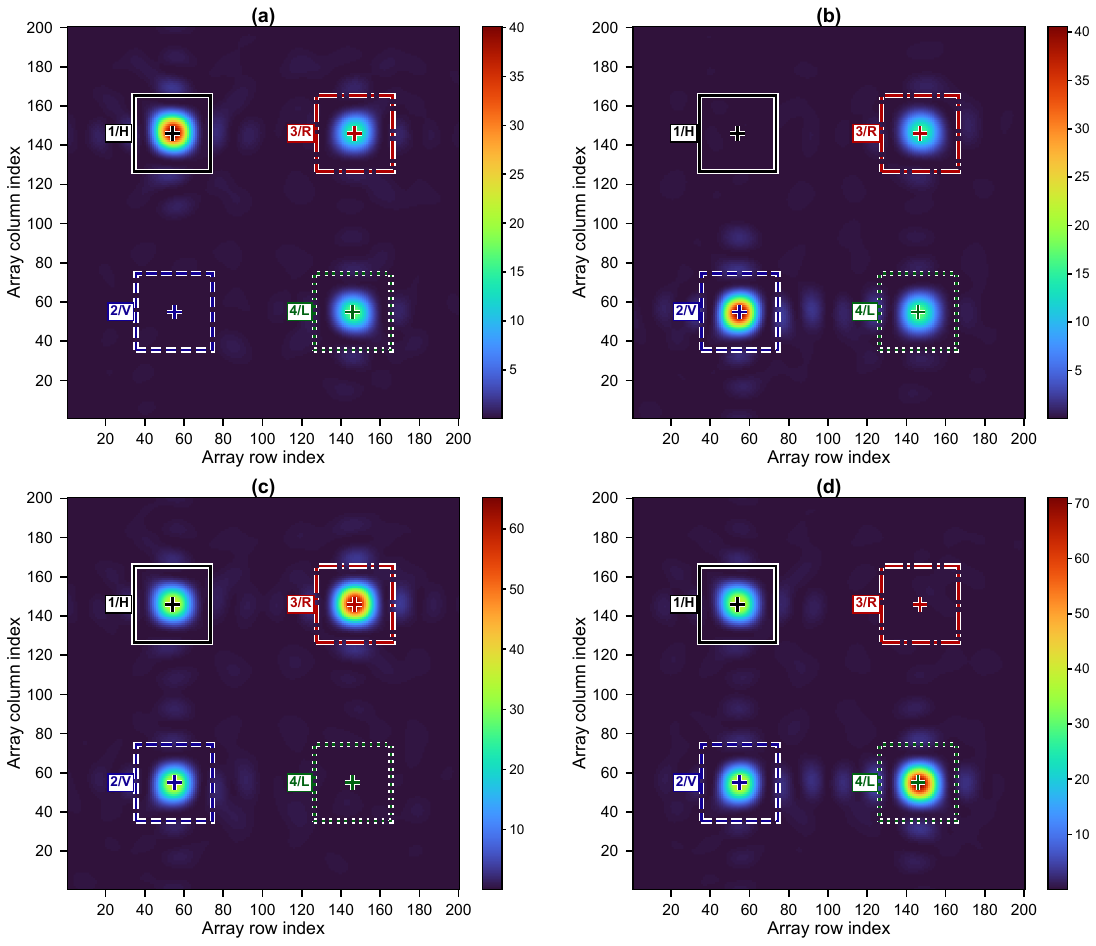}
\caption{Representative integration-region overlay for the selected 1550~nm P750 device under \textbf{(a)} $H$, \textbf{(b)} $V$, \textbf{(c)} $R$, and \textbf{(d)} $L$ input. The raw $200\times200$ intensity maps are displayed against the simulation row and column indices, and the final $39\times39$-pixel detector boundaries are drawn explicitly on every input field.}
\label{fig:integrationaudit}
\end{figure}

All selected ports retain more than 99\% of the fitted core and at least 95\% of the empirical background-corrected reference signal. Integration boundaries were visually verified on every raw intensity map; a representative overlay is shown in \Cref{fig:integrationaudit}.

For input $i$, the final raw sums $S_{i,k}=\sum_{\Omega_k}I_i$ are normalized over the four calibrated regions. The port assignment found in the first pass is used unchanged for the four states of the candidate.

The fraction of the sampled focal plane captured by the four regions is
\begin{equation}
f_i^{\rm ROI}=\frac{\sum_{k=1}^4S_{i,k}}{\sum_{r,c}I_i(r,c)},
\end{equation}
and the collected efficiency is $\eta_i^{\rm coll}=|T_i|f_i^{\rm ROI}$. The absolute value retains the power magnitude when a monitor sign is set by its normal-vector convention.

\section{Complete selected response matrices}

The matrices below are printed to six decimal places.

\begin{table}[h]
\centering
\caption{Selected 780~nm Si P380 conditional response.}
\begin{tabular}{@{}lrrrr@{}}
\toprule
Input & H & V & R & L \\
\midrule
H & 0.450850 & 0.030616 & 0.258921 & 0.259613 \\
V & 0.026352 & 0.447040 & 0.253107 & 0.273501 \\
R & 0.225177 & 0.257209 & 0.506802 & 0.010813 \\
L & 0.229640 & 0.242504 & 0.009662 & 0.518195 \\
\bottomrule
\end{tabular}
\end{table}

\begin{table}[h]
\centering
\caption{Selected 1550~nm Si P750 conditional response.}
\begin{tabular}{@{}lrrrr@{}}
\toprule
Input & H & V & R & L \\
\midrule
H & 0.531417 & 0.005323 & 0.222774 & 0.240486 \\
V & 0.008201 & 0.503528 & 0.234781 & 0.253490 \\
R & 0.268919 & 0.253883 & 0.467320 & 0.009878 \\
L & 0.264643 & 0.260620 & 0.009903 & 0.464834 \\
\bottomrule
\end{tabular}
\end{table}

\begin{table}[h]
\centering
\caption{Selected 2000~nm Si P900 conditional response.}
\begin{tabular}{@{}lrrrr@{}}
\toprule
Input & H & V & R & L \\
\midrule
H & 0.562943 & 0.010605 & 0.209840 & 0.216612 \\
V & 0.016399 & 0.572888 & 0.213732 & 0.196981 \\
R & 0.301038 & 0.281407 & 0.410083 & 0.007471 \\
L & 0.303630 & 0.275976 & 0.006802 & 0.413592 \\
\bottomrule
\end{tabular}
\end{table}

\begin{table}[h]
\centering
\caption{Selected 10.6~$\mu$m Si conditional response.}
\begin{tabular}{@{}lrrrr@{}}
\toprule
Input & H & V & R & L \\
\midrule
H & 0.446710 & 0.007374 & 0.267663 & 0.278252 \\
V & 0.006884 & 0.409588 & 0.289802 & 0.293726 \\
R & 0.219998 & 0.207363 & 0.565209 & 0.007431 \\
L & 0.226292 & 0.215901 & 0.006052 & 0.551755 \\
\bottomrule
\end{tabular}
\end{table}

\section{Metrics and complete candidate comparison}

For probability distributions $C_i$ and $C_i^{\rm ideal}$, the state fidelity and total-variation distance are
\begin{equation}
F_i=\left(\sum_m\sqrt{C_{i,m}C^{\rm ideal}_{i,m}}\right)^2,
\quad D_i^{\rm TV}=\tfrac12\sum_m|C_{i,m}-C^{\rm ideal}_{i,m}|.
\end{equation}
The basis errors are
\begin{equation}
Q_{HV}=\frac{C_{H,V}+C_{V,H}}{C_{H,H}+C_{H,V}+C_{V,H}+C_{V,V}},
\end{equation}
with an analogous $Q_{RL}$. \Cref{tab:candidates} lists all six complete candidates.

\begin{table}[h]
\centering
\caption{Complete candidate comparison under the four-state optical benchmark. Asterisks mark the selected design at each wavelength. The characterized-receiver security yields are reported separately in \Cref{tab:si_security_yields}.}
\label{tab:candidates}
\scriptsize
\begin{tabular}{@{}lrrrrrrr@{}}
\toprule
Candidate & $\overline Q$ & $\overline F$ & $|T|$ & $f^{\rm ROI}$ & $\eta^{\rm coll}$ & $r_{\rm sec}$ & $Y_{\rm sec}$ \\
\midrule
780 Si P380$^*$ & 0.0396 & 0.9796 & 0.3755 & 0.6078 & 0.2282 & 0.4826 & 0.0551 \\
780 TiO$_2$ P360 & 0.0559 & 0.9215 & 0.4956 & 0.6357 & 0.3150 & 0.3274 & 0.0516 \\
1550 Si P700 & 0.0252 & 0.9865 & 0.5547 & 0.8388 & 0.4653 & 0.6324 & 0.1471 \\
1550 Si P750$^*$ & 0.0168 & 0.9908 & 0.5363 & 0.7634 & 0.4094 & 0.7388 & 0.1513 \\
2000 Si P900$^*$ & 0.0201 & 0.9830 & 0.4801 & 0.7194 & 0.3454 & 0.6909 & 0.1193 \\
10.6 $\mu$m Si$^*$ & 0.0142 & 0.9885 & 0.3020 & 0.5771 & 0.1743 & 0.7668 & 0.0668 \\
\bottomrule
\end{tabular}
\end{table}

\begin{figure}[!htbp]
\centering
\includegraphics[width=0.92\textwidth]{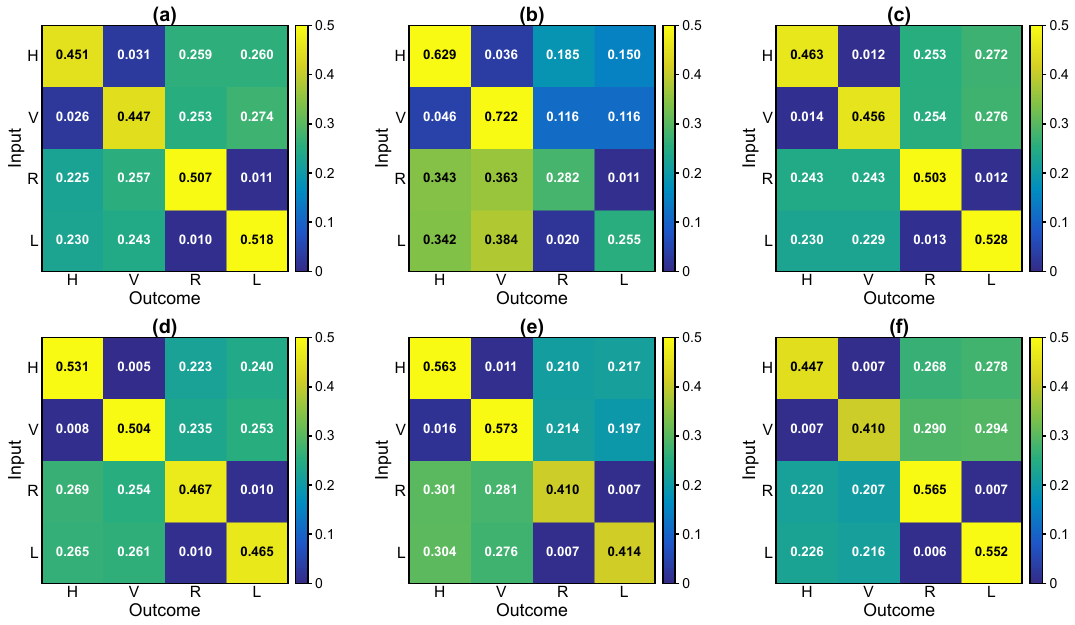}
\caption{Response-matrix comparison for all complete candidates. \textbf{(a)} 780~nm Si P380, selected, with 3.96\% mean QBER. \textbf{(b)} 780~nm TiO$_2$ P360, with 5.59\% mean QBER. \textbf{(c)} 1550~nm Si P700, with 2.52\% mean QBER. \textbf{(d)} 1550~nm Si P750, selected, with 1.68\% mean QBER. \textbf{(e)} 2~$\mu$m Si P900, selected, with 2.01\% mean QBER. \textbf{(f)} 10.6~$\mu$m Si P4700, selected, with 1.42\% mean QBER.}
\label{fig:candidates}
\end{figure}

Under this optical benchmark, the P700 design collects more power than P750 (46.53\% versus 40.94\%), but P750 has the lower mean QBER (1.68\% versus 2.52\%) and the larger device-level secret yield (0.1513 versus 0.1471 bits per incident photon). P750 is therefore selected. Similarly, the 780~nm TiO$_2$ candidate has higher collection than Si P380, but its response is less faithful to the target and its mean QBER is larger. Propagating both effects gives a yield of 0.0516 for TiO$_2$ and 0.0551 for Si P380, so the silicon design is selected.

\section{Detector-region sensitivity}

The four calibration-defined detector squares are simultaneously contracted or expanded by margins $-5,-2,0,+2,+5$ pixels. Negative margins isolate the bright spot cores; positive margins collect more of each halo. The fitted centres and port labels remain fixed throughout this test.

\begin{figure}[!htbp]
\centering
\includegraphics[width=0.92\textwidth]{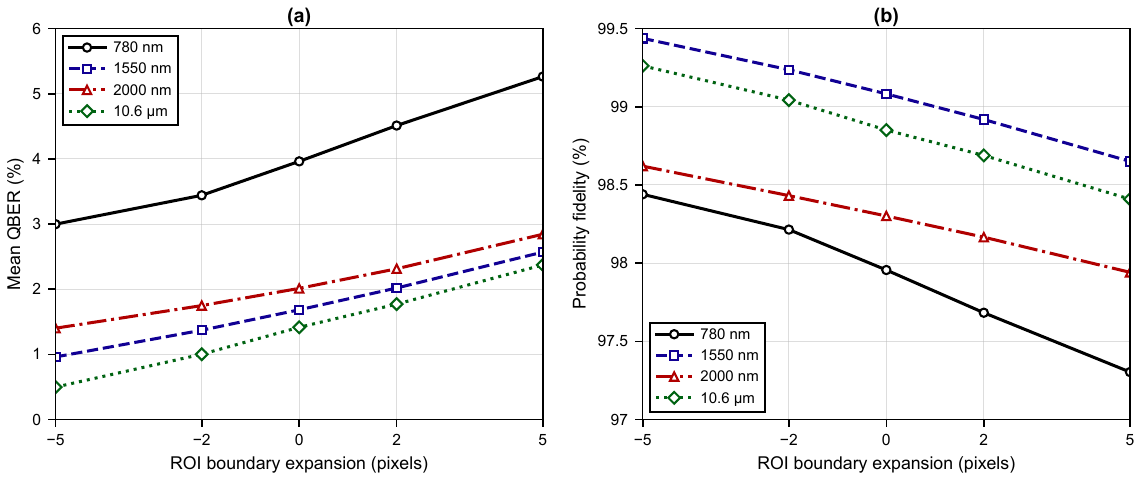}
\caption{Detector-region sensitivity. \textbf{(a)} Mean QBER and \textbf{(b)} probability fidelity versus detector-region boundary expansion.}
\label{fig:roisensitivity}
\end{figure}

\begin{table}[h]
\centering
\caption{Range over $-5$ to $+5$ pixel boundary changes. The nominal value is shown in parentheses.}
\small
\begin{tabular}{@{}lccc@{}}
\toprule
$\lambda$ & Mean QBER & $\overline F$ & $Y_{\rm sec}$ \\
\midrule
780 nm & 0.0300--0.0526 (0.0396) & 0.9730--0.9844 & 0.0456--0.0595 \\
1550 nm & 0.0096--0.0257 (0.0168) & 0.9865--0.9944 & 0.1398--0.1655 \\
2000 nm & 0.0140--0.0284 (0.0201) & 0.9794--0.9862 & 0.1124--0.1257 \\
10.6 $\mu$m & 0.0050--0.0237 (0.0142) & 0.9841--0.9926 & 0.0590--0.0684 \\
\bottomrule
\end{tabular}
\end{table}

The detector aperture is therefore part of the receiver design. A larger aperture is not automatically beneficial here since, while it increases collection, it also can accept diffuse power from the orthogonal matched-basis channel. A physical implementation should optimize detector/fibre mode overlap jointly with the security metric.

\begin{figure}[!htbp]
\centering
\includegraphics[width=0.88\textwidth]{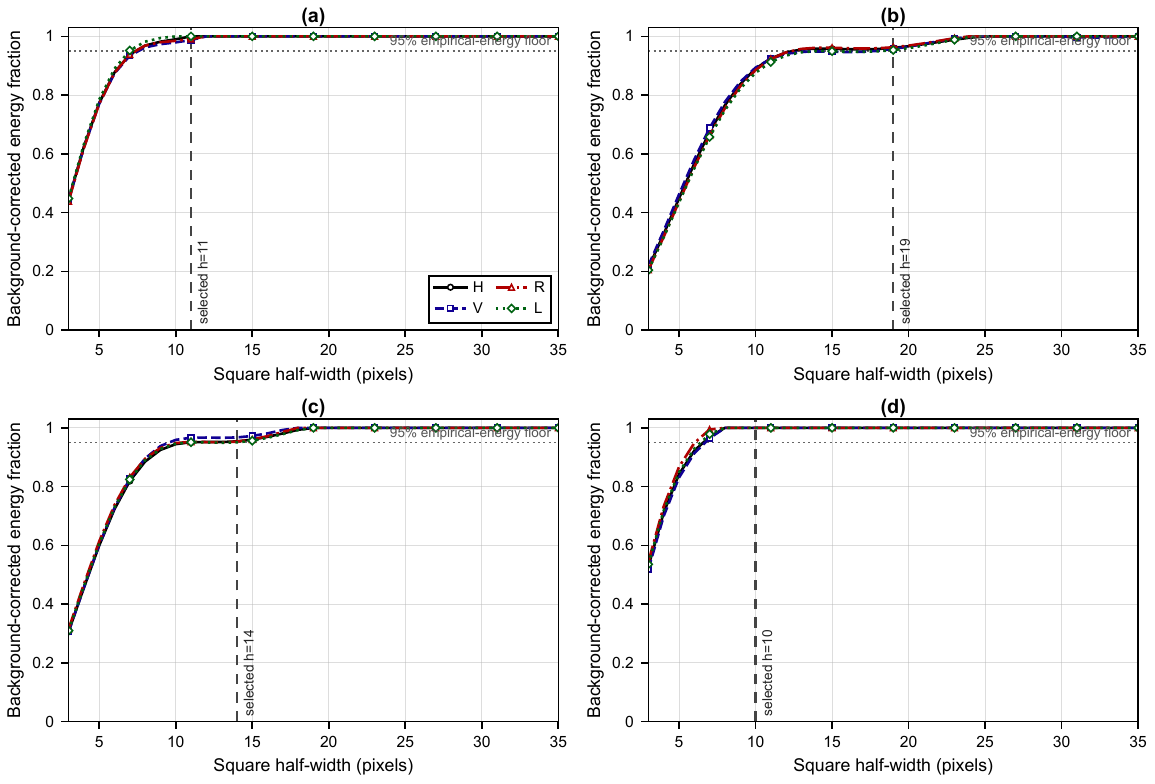}
\caption{Quantitative aperture selection for the chosen designs at \textbf{(a)} 780~nm, \textbf{(b)} 1550~nm, \textbf{(c)} 2000~nm, and \textbf{(d)} 10.6~$\mu$m. The selected detector-square sides are 23, 39, 29, and 21 pixels, respectively, and the corresponding largest fitted-baseline fractions are 4.0\%, 4.9\%, 3.7\%, and 4.9\%. Curves show the empirical background-corrected signal fraction for each logical port relative to the 35-pixel-half-width reference. The selected vertical line is the smallest common half-width that also retains at least 99\% of every fitted Gaussian core and limits every estimated constant-baseline contribution to 10\%.}
\label{fig:apertureselection}
\end{figure}

\section{Four-state optical benchmark and added-error test}

For a transparent optical comparison, the single-photon asymptotic secret fraction per sifted click is
\begin{equation}
r_{\rm sec}=\max\{0,1-f_{\rm EC}\binaryentropy(Q_{HV})-\binaryentropy(Q_{RL})\},
\label{eq:skey}
\end{equation}
with $f_{\rm EC}=1.16$ and $\binaryentropy(x)=-x\log_2x-(1-x)\log_2(1-x)$ \cite{SI-ShorPreskill2000}. The yield per incident photon is $Y_{\rm sec}=\overline p_{\rm sift}r_{\rm sec}$, where $\overline p_{\rm sift}$ averages the absolute matching-outcome probabilities over the four equally likely input states. Transmission, collection and the receiver's basis-selection weight are already included in those probabilities; no additional passive-basis factor is applied.

For the added-error curves in the main paper, a bit-flip probability $e_{\rm add}$ is applied symmetrically within each basis. If $Q$ is the nominal basis error, the updated value is
\begin{equation}
Q'=Q(1-e_{\rm add})+(1-Q)e_{\rm add}.
\end{equation}

The optical benchmark is evaluated at the device plane with ideal single-photon inputs. It omits:
\begin{itemize}
\item channel attenuation and polarization drift;
\item detector quantum efficiency, dark counts, dead time, afterpulsing, and timing windows;
\item source vacuum/multiphoton statistics and decoy-state estimation;
\item finite-size parameter estimation and composable-security terms;
\item arbitrary spatial/spectral input modes outside the admitted qubit mode.
\end{itemize}
The yields are expressed per photon incident on the device. Section~\ref{sec:si_security_v4} extends this comparison to the reconstructed receiver operators, including their polarization-dependent collection and nonideal basis selection.

\section{10.6~$\mu$m numerical model}

The 10.6~$\mu$m model uses a 4.7~$\mu$m lattice period, a 117.5~$\mu$m aperture radius, a 150~$\mu$m focal length, and focal offsets of $\pm60~\mu$m. The four input-polarization fields are processed using the same detector-region calibration, port assignment, and response-matrix analysis as the near-infrared designs.

\FloatBarrier
\clearpage
\section{Security with the reconstructed measurement operators}
\label{sec:si_security_v4}

The six-state simulations characterize the receiver as a four-outcome qubit measurement with an additional no-click outcome. The security calculation uses the reconstructed operators, including unequal collection, wrong-port leakage and state-dependent basis selection. H/V and R/L matched outcomes both contribute to the secret key.

\subsection{Ideal basis independence}

Let $\hat E_H,\hat E_V,\hat E_R,\hat E_L$ denote the four click operators and let $\hat F=\sum_j\hat E_j$. State-independent selection of the two bases is expressed by
\begin{equation}
\hat E_H+\hat E_V=p\hat F,\qquad
\hat E_R+\hat E_L=(1-p)\hat F.
\label{eq:si_basis_independent}
\end{equation}
For every incident polarization state $\rho$ with nonzero click probability, Eq.~\eqref{eq:si_basis_independent} gives $P(H/V\mid{\rm click},\rho)=p$. It also holds when the incident photon is correlated with Eve. If $\rho_{BE}$ is their joint state, the unnormalized conditional states held by Eve satisfy
\begin{align}
\widetilde\rho_E^{\,Z}
 &=\operatorname{Tr}_B[(\hat S_Z\otimes\hat I_E)\rho_{BE}]
 =p\,\widetilde\rho_E^{\,{\rm click}},\\
\widetilde\rho_E^{\,Y}
 &=(1-p)\,\widetilde\rho_E^{\,{\rm click}},
\end{align}
where $\hat S_Z=\hat E_H+\hat E_V$ and $\hat S_Y=\hat E_R+\hat E_L$. Their normalized states are identical. Thus the receiver's basis choice carries no information to Eve before public basis announcement. A known constant 49:51 split satisfies this condition just as a 50:50 split does. Complementary-basis correlations determine the secrecy of the retained bit values, as derived below.

\subsection{Six-state reconstruction}

The probe states are $H,V,R,L$ together with
\begin{equation}
\ket D=(\ket H+\ket V)/\sqrt2,\qquad
\ket A=(\ket H-\ket V)/\sqrt2.
\end{equation}
H/V, R/L and D/A are the Pauli-$Z$, Pauli-$Y$ and Pauli-$X$ eigenstates, respectively, with the circular convention defined in Section~2. The integrated probabilities per incident photon are
\begin{equation}
q_{j|s}=\eta_{\rm coll}(s)C_{j|s},\qquad
q_{\emptyset|s}=1-\sum_{j=H,V,R,L}q_{j|s}.
\label{eq:si_absolute}
\end{equation}
The same four physical detector regions are applied to every input. The raw intensities, integrations and complete probability tables are given in Section~\ref{sec:si_DA_data}.

Each detector operator has the form
\begin{equation}
\hat E_j=\frac12\bigl(t_j\hat I+x_j\hat\sigma_x+y_j\hat\sigma_y+z_j\hat\sigma_z\bigr)
=\frac12
\begin{pmatrix}
t_j+z_j&x_j-i y_j\\
x_j+i y_j&t_j-z_j
\end{pmatrix}.
\label{eq:si_operator_matrix}
\end{equation}
Exactly consistent probabilities obey
\begin{align}
t_j&=q_{j|H}+q_{j|V}=q_{j|R}+q_{j|L}=q_{j|D}+q_{j|A},\\
x_j&=q_{j|D}-q_{j|A},\qquad
y_j=q_{j|R}-q_{j|L},\qquad
z_j=q_{j|H}-q_{j|V}.
\end{align}
The unconstrained six-state inversion averages the three estimates of $t_j$. A constrained least-squares fit then imposes
\begin{equation}
\hat E_j\succeq0,\qquad
\hat E_{\emptyset}=\hat I-\sum_j\hat E_j\succeq0.
\end{equation}
D/A therefore supplies the missing $x_j$ coefficient, separate from the key basis. The full-precision physical operators and their nominal predictions are used in the security calculation. The raw probabilities and reconstruction residuals are reported separately.

\subsection{Polarization dependence of basis selection}

For a characterized receiver, the eigenvalues of
\begin{equation}
\hat B=\hat F^{-1/2}\hat S_Z\hat F^{-1/2}
\end{equation}
bound the basis probability over all polarization states:
\begin{equation}
b_{\min}\leq P(H/V\mid{\rm click},\rho)\leq b_{\max}.
\end{equation}
The overall deviation from 50:50 is
$\epsilon=\max(|b_{\min}-1/2|,|b_{\max}-1/2|)$.
The half-range $(b_{\max}-b_{\min})/2$ isolates the state dependence from a constant basis preference. These quantities are evaluated on the support of $\hat F$; the reconstructed models here have full-rank acceptance operators.

\begin{table}[!htbp]
\caption{Basis-selection bounds for the six reconstructed receivers. The half-range quantifies polarization dependence; $\epsilon$ also includes a constant preference for one basis.}
\label{tab:si_basis_bounds}

\centering
\small
\begin{tabular}{@{}lrrrr@{}}
\toprule
Fixed receiver & $b_{\min}$ & $b_{\max}$ & $\epsilon=\max|b-\tfrac12|$ & Half-range \\
\midrule
1550 nm Si P750 & 0.511452 & 0.536766 & 0.036766 & 0.012657 \\
1550 nm Si P700 & 0.458631 & 0.486405 & 0.041369 & 0.013887 \\
780 nm Si P380 & 0.491390 & 0.506469 & 0.008610 & 0.007540 \\
780 nm TiO$_2$ P360 & 0.664695 & 0.769599 & 0.269599 & 0.052452 \\
2000 nm Si P900 & 0.573241 & 0.589378 & 0.089378 & 0.008068 \\
10.6 $\mu$m Si P4700 & 0.414730 & 0.455383 & 0.085270 & 0.020326 \\
\bottomrule
\end{tabular}

\end{table}

\subsection{Both-basis secret-key bound}
\label{sec:si_both_basis_proof}

The protocol uses ideal single-photon preparations in the admitted optical mode, a fixed trusted qubit receiver including no-click, and authenticated public communication. The bound is asymptotic against collective attacks. Alice chooses H/V and R/L with equal probability. Random disclosed subsets estimate the correlations; the undisclosed matched outcomes from both bases supply key bits. All conditional entropies below include the public basis label and the accepted-event announcement. The error-correction inefficiency is $f_{\rm EC}=1.16$.

For key basis $b\in\{Z,Y\}$, define
\begin{align}
\hat S_Z&=\hat E_H+\hat E_V,&
\hat W_Z&=\hat E_R-\hat E_L,\\
\hat S_Y&=\hat E_R+\hat E_L,&
\hat W_Y&=\hat E_H-\hat E_V,\\
\kappa_b&=\left\|\hat S_b^{-1/2}\hat W_b\hat S_b^{-1/2}\right\|_\infty.
\end{align}
Here $\hat S_b$ selects accepted key events, while $\hat W_b$ measures the complementary correlation. From the operating probabilities,
\begin{align}
p_Z&=\tfrac12(q_{H|H}+q_{V|H}+q_{H|V}+q_{V|V}),\\
p_Y&=\tfrac12(q_{R|R}+q_{L|R}+q_{R|L}+q_{L|L}),\\
e_Z&=\frac{q_{V|H}+q_{H|V}}{2p_Z},\qquad
e_Y=\frac{q_{L|R}+q_{R|L}}{2p_Y},\\
\tau_Y&=\tfrac12(q_{R|R}+q_{L|L}-q_{L|R}-q_{R|L}),\\
\tau_Z&=\tfrac12(q_{H|H}+q_{V|V}-q_{V|H}-q_{H|V}).
\end{align}
Thus $p_b$ is the matching-click probability given Alice's basis, $e_b$ is its bit error, and $\tau_b$ is an unconditioned signed correlation.

In the entanglement-based representation, acceptance in basis $b$ is a virtual filter $\sqrt{\hat S_b}$. On the accepted ensemble,
\begin{equation}
\hat O_b=\hat S_b^{-1/2}\hat W_b\hat S_b^{-1/2}/\kappa_b
\end{equation}
is a valid binary observable because $-\hat I\preceq\hat O_b\preceq\hat I$, equivalently
$\hat S_b\pm\hat W_b/\kappa_b\succeq0$.
The complementary prediction errors are therefore
\begin{equation}
e_{\rm ph,Z}=\frac12\left(1-\frac{|\tau_Y|}{\kappa_Zp_Z}\right),\qquad
e_{\rm ph,Y}=\frac12\left(1-\frac{|\tau_Z|}{\kappa_Yp_Y}\right).
\label{eq:si_phase_error}
\end{equation}
The transpose of the circular source observable changes its sign in the entanglement-based representation; the absolute correlation accounts for this bit relabelling. State-dependent acceptance is included through $\hat S_b$ and $\kappa_b$, so Eq.~\eqref{eq:si_phase_error} does not replace a virtual phase error by an uncorrected opposite-basis QBER.

The uncertainty relation with quantum memory and the binary prediction-error bound give~\cite{SI-Berta2010}
\begin{equation}
H(A_b\mid E,{\rm pass}_b)\geq1-h_2(e_{\rm ph,b}).
\end{equation}
Eve's Holevo information before reconciliation is correspondingly at most $h_2(e_{\rm ph,b})$ bits per retained bit. After accounting for reconciliation, the one-way secret fraction is~\cite{SI-DevetakWinter2005}
\begin{equation}
r_b=\bigl[1-h_2(e_{\rm ph,b})-1.16h_2(e_b)\bigr]_+,
\end{equation}
where $[u]_+=\max(0,u)$. Privacy amplification is applied to the two basis-labelled substrings. Their secret lengths add:
\begin{equation}
Y_{\rm tom}=\frac12p_Zr_Z+\frac12p_Yr_Y,\qquad
r_{\rm tom}=\frac{p_Zr_Z+p_Yr_Y}{p_Z+p_Y}.
\label{eq:si_total_yield}
\end{equation}
$Y_{\rm tom}$ is secret bits per photon incident on the device, and $r_{\rm tom}$ is secret bits per sifted bit. For an ideal receiver of efficiency $\eta$, Eq.~\eqref{eq:si_total_yield} gives $\eta/2$, including a known fixed basis bias, without assuming exact equality of the basis-selection probabilities.

\subsection{Characterized-receiver yields}

All six reconstructed models give a positive total yield (Table~\ref{tab:si_security_yields}). The P750 result is $0.15164$ secret bits per incident photon, close to its four-state optical benchmark of $0.15126$. P700 gives $0.15935$ under the characterized-receiver bound, the largest value in this six-device set. The benchmark selection and its figures remain distinct from this operator-based comparison.

\begin{table}[!htbp]
\caption{Secret fractions and yields for the reconstructed receivers. $r_{\rm HV}$ and $r_{\rm RL}$ are per retained bit in their respective bases; $Y_{\rm HV}$ and $Y_{\rm RL}$ include Alice's basis probability and are per incident photon. Their sum is $Y_{\rm tom}$.}
\label{tab:si_security_yields}

\centering
\small
\begin{tabular}{@{}lrrrrr@{}}
\toprule
Fitted model & $r_{\rm HV}$ & $r_{\rm RL}$ & $Y_{\rm HV}$ & $Y_{\rm RL}$ & $Y_{\rm total}$ \\
\midrule
1550 nm Si P750 & 0.81185 & 0.65548 & 0.08750 & 0.06414 & 0.15164 \\
1550 nm Si P700 & 0.61668 & 0.74612 & 0.06776 & 0.09159 & 0.15935 \\
780 nm Si P380 & 0.38431 & 0.77223 & 0.02285 & 0.04616 & 0.06901 \\
780 nm TiO$_2$ P360 & 0.38185 & 0.12521 & 0.04289 & 0.00558 & 0.04846 \\
2000 nm Si P900 & 0.78224 & 0.79457 & 0.07851 & 0.05749 & 0.13599 \\
10.6 $\mu$m Si P4700 & 0.75290 & 0.66511 & 0.02855 & 0.03276 & 0.06130 \\
\bottomrule
\end{tabular}

\end{table}

\begin{table}[!htbp]
\caption{Acceptance, bit errors and complementary-prediction quantities entering each basis contribution. The definitions are those of Section~\ref{sec:si_both_basis_proof}.}
\label{tab:si_basis_calculation}

\centering
\small
\begin{tabular}{@{}llrrrrr@{}}
\toprule
Fixed receiver & Basis & $p_b$ & $e_b$ & $\kappa_b$ & $e_{{\rm ph},b}$ & $r_b$ \\
\midrule
1550 nm Si P750 & H/V & 0.21556 & 0.01504 & 0.87794 & 0.00663 & 0.81185 \\
1550 nm Si P750 & R/L & 0.19571 & 0.02294 & 1.12141 & 0.02369 & 0.65548 \\
1550 nm Si P700 & H/V & 0.21974 & 0.02678 & 1.12410 & 0.02656 & 0.61668 \\
1550 nm Si P700 & R/L & 0.24552 & 0.02369 & 0.86061 & 0.00787 & 0.74612 \\
780 nm Si P380 & H/V & 0.11890 & 0.09654 & 0.98504 & 0.01056 & 0.38431 \\
780 nm Si P380 & R/L & 0.11954 & 0.02047 & 0.81412 & 0.00707 & 0.77223 \\
780 nm TiO$_2$ P360 & H/V & 0.22462 & 0.05586 & 0.38875 & 0.04339 & 0.38185 \\
780 nm TiO$_2$ P360 & R/L & 0.08909 & 0.05248 & 2.94970 & 0.12039 & 0.12521 \\
2000 nm Si P900 & H/V & 0.20072 & 0.02304 & 0.70144 & 0.00358 & 0.78224 \\
2000 nm Si P900 & R/L & 0.14470 & 0.01698 & 1.34248 & 0.00718 & 0.79457 \\
10.6 $\mu$m Si P4700 & H/V & 0.07583 & 0.01635 & 1.30516 & 0.01419 & 0.75290 \\
10.6 $\mu$m Si P4700 & R/L & 0.09850 & 0.01187 & 0.80370 & 0.03672 & 0.66511 \\
\bottomrule
\end{tabular}

\end{table}

The bound includes the nonideal polarization response and loss of each reconstructed receiver. The figures represent the nominal simulated operating point. Channel and detector factors can be applied only within a correspondingly characterized source-and-receiver model.

\subsection{Numerical consistency and operator uncertainty}
\label{sec:si_consistency}

The opposite-state sums test whether the six raw responses admit one exact qubit POVM. Table~\ref{tab:si_reconstruction_residuals} reports their discrepancies and the residual of the physical reconstruction. The quoted nominal yields are conditional bounds for the fitted receiver models and their operating statistics. The numerical residual describes agreement with the simulations.

For a specified operator-norm allowance
$\|\hat E_j-\hat E_j^0\|_\infty\leq\delta$,
a robust virtual predictor is obtained by choosing the largest $c_b\geq0$ satisfying
\begin{equation}
\hat S_b^0\pm c_b\hat W_b^0\succeq2\delta(1+c_b)\hat I .
\label{eq:si_robust_operator}
\end{equation}
Then $\hat S_b\pm c_b\hat W_b\succeq0$ for every receiver in this operator set. Replacing $1/\kappa_b$ by $c_b$ in Eq.~\eqref{eq:si_phase_error} and summing both basis contributions gives the sensitivity bounds in Table~\ref{tab:si_delta}. The operating statistics are held fixed. At $\delta=0.005$, all six models retain a positive bound; P750 gives $0.10639$ secret bits per incident photon. Here $\delta$ is the stated uncertainty-set radius. Finite-key terms and experimental calibration uncertainties are outside the stated asymptotic simulation model.

\begin{table}[!htbp]
\caption{Both-basis secret yield for specified per-operator norm allowances $\delta$, with nominal operating statistics held fixed. All values are secret bits per incident photon.}
\label{tab:si_delta}

\centering
\small
\begin{tabular}{@{}lrrrr@{}}
\toprule
Original receiver & $\delta=0$ & $\delta=0.001$ & $\delta=0.005$ & $\delta=0.010$ \\
\midrule
1550 nm Si P750 & 0.15164 & 0.14033 & 0.10639 & 0.07638 \\
1550 nm Si P700 & 0.15935 & 0.14804 & 0.11318 & 0.08154 \\
780 nm Si P380 & 0.06901 & 0.05660 & 0.02524 & 0.01093 \\
780 nm TiO$_2$ P360 & 0.04846 & 0.03981 & 0.01574 & 0.00000 \\
2000 nm Si P900 & 0.13599 & 0.12217 & 0.08527 & 0.05530 \\
10.6 $\mu$m Si P4700 & 0.06130 & 0.05205 & 0.02832 & 0.01181 \\
\bottomrule
\end{tabular}

\end{table}
\FloatBarrier

\section{D/A focal fields and complete tomography data}
\label{sec:si_DA_data}

\subsection{Spot calibration and physical detector regions}

All twelve D/A focal fields are sampled on their native $200\times200$ grids. Each of the 48 D/A spots is fitted with the elliptical Gaussian plus constant baseline of Eq.~\eqref{eq:sfit}, using a 61-pixel search window, the soft-$L_1$ residual and the Nelder--Mead optimizer. Native-grid calibration uses a common square half-width $h$ for all four ports of each input, chosen from $3,\ldots,35$ pixels. The criteria are 99\% fitted-core retention, 95\% net empirical signal relative to a 71-pixel reference square and at most 10\% estimated baseline. All twelve D/A calibrations meet these criteria.

The detector geometry used in tomography remains the original four physical squares of each device, fixed for all six inputs. D/A fits and growth curves verify the spot centres and collection without allowing the receiver to change with the input state. Every final integral uses the raw intensity, including the baseline.

The spatial grid differs between the original and D/A exports for two devices: P750 spans $\pm22.5~\mu$m and $\pm21.3~\mu$m, respectively; TiO$_2$ P360 spans $\pm9.1~\mu$m and $\pm9.6~\mu$m. The original detector boundaries are therefore applied by physical pixel-area overlap. For $w_{uv}^{(j)}$ equal to the area of native pixel $(u,v)$ inside detector $j$,
\begin{align}
J_j(s)&=\sum_{u,v}w_{uv}^{(j)}I_s(u,v),&
C_{j|s}&=\frac{J_j(s)}{\sum_kJ_k(s)},\\
q_{j|s}&=|T_s|\frac{J_j(s)}{J_{\rm plane}(s)}.
\label{eq:si_physical_integration}
\end{align}
Here $|T_s|$ is transmitted power divided by incident power, and $J_{\rm plane}$ integrates the full exported focal plane. This is the same sampled-plane normalization as the four-state analysis. No interpolation of the simulated field or renormalization of one probe basis relative to another is applied.

\begin{table}[!htbp]
\caption{D/A conditional responses with the original fixed physical detector regions. The ideal values are $C_H=C_V=C_R=C_L=1/4$. The collection efficiency $\eta_{\rm coll}=\sum_jq_{j|s}$ is per incident photon.}
\label{tab:si_da_response}

\centering
\small
\begin{tabular}{@{}llrrrrr@{}}
\toprule
Device & Input & $C_H$ & $C_V$ & $C_R$ & $C_L$ & $\eta_{\rm coll}$ \\
\midrule
1550 nm Si P750 & D & 0.26487 & 0.25732 & 0.24383 & 0.23398 & 0.41820 \\
1550 nm Si P750 & A & 0.26853 & 0.25747 & 0.21354 & 0.26046 & 0.41196 \\
1550 nm Si P700 & D & 0.23425 & 0.23524 & 0.24163 & 0.28889 & 0.46930 \\
1550 nm Si P700 & A & 0.23893 & 0.23623 & 0.26638 & 0.25846 & 0.46123 \\
780 nm Si P380 & D & 0.25493 & 0.27857 & 0.21865 & 0.24785 & 0.26103 \\
780 nm Si P380 & A & 0.27254 & 0.26685 & 0.24147 & 0.21914 & 0.25678 \\
780 nm TiO$_2$ P360 & D & 0.33532 & 0.37676 & 0.15714 & 0.13077 & 0.30753 \\
780 nm TiO$_2$ P360 & A & 0.34960 & 0.37047 & 0.14481 & 0.13512 & 0.31464 \\
2000 nm Si P900 & D & 0.28894 & 0.29320 & 0.21348 & 0.20438 & 0.34744 \\
2000 nm Si P900 & A & 0.31580 & 0.26415 & 0.20989 & 0.21016 & 0.34342 \\
10.6 $\mu$m Si P4700 & D & 0.22953 & 0.20752 & 0.28138 & 0.28156 & 0.17522 \\
10.6 $\mu$m Si P4700 & A & 0.21681 & 0.21604 & 0.27641 & 0.29073 & 0.17344 \\
\bottomrule
\end{tabular}

\end{table}

\subsection{Consistency of the six-state probabilities}

For each detector, the three opposite-state sums in the tomography equations are equal for an exact fixed qubit receiver. $\Delta_{\rm pair}$ below is their largest pairwise disagreement; the maximum residual and root-mean-square error (RMSE) compare the raw click probabilities with the positive-operator reconstruction.

\begin{table}[!htbp]
\caption{Numerical consistency and physical-reconstruction residuals. All entries are absolute probabilities per incident photon.}
\label{tab:si_reconstruction_residuals}

\centering
\small
\begin{tabular}{@{}lrrr@{}}
\toprule
Device & $\Delta_{\rm pair}$ & Max. residual & RMSE \\
\midrule
1550 nm Si P750 & 0.007301 & 0.001944 & 0.001100 \\
1550 nm Si P700 & 0.000978 & 0.000303 & 0.000166 \\
780 nm Si P380 & 0.032710 & 0.010800 & 0.005019 \\
780 nm TiO$_2$ P360 & 0.003980 & 0.001111 & 0.000548 \\
2000 nm Si P900 & 0.002824 & 0.000938 & 0.000467 \\
10.6 $\mu$m Si P4700 & 0.000188 & 0.000062 & 0.000033 \\
\bottomrule
\end{tabular}

\end{table}

The Si P380 D/A total-transmission sum is 13.6\% above its H/V sum before detector integration. The other devices have smaller detector-level pair-sum discrepancies, as shown in Table~\ref{tab:si_reconstruction_residuals}. All six fitted POVMs have positive click and no-click operators. The residuals are retained as numerical consistency diagnostics, separately from the specified operator uncertainty in Eq.~\eqref{eq:si_robust_operator}.

\FloatBarrier
\subsection{Complete input probabilities}

The following tables report raw integrated probabilities per incident photon for H,V,R,L,D,A. They include the no-click probability, so every row sums to one up to rounding.

\noindent\begin{minipage}{\textwidth}
\centering
{\bfseries 1550 nm Si P750\par}

\centering
\footnotesize
\begin{tabular}{@{}lrrrrr@{}}
\toprule
Input & $q_H$ & $q_V$ & $q_R$ & $q_L$ & $q_{\emptyset}$ \\
\midrule
H & 0.217717 & 0.002181 & 0.091268 & 0.098525 & 0.590309 \\
V & 0.003355 & 0.205971 & 0.096039 & 0.103691 & 0.590944 \\
R & 0.108004 & 0.101965 & 0.187686 & 0.003967 & 0.598377 \\
L & 0.110384 & 0.108706 & 0.004131 & 0.193884 & 0.582896 \\
D & 0.110772 & 0.107611 & 0.101969 & 0.097853 & 0.581795 \\
A & 0.110624 & 0.106069 & 0.087971 & 0.107299 & 0.588036 \\
\bottomrule
\end{tabular}
\end{minipage}\par\medskip

\noindent\begin{minipage}{\textwidth}
\centering
{\bfseries 1550 nm Si P700\par}

\centering
\footnotesize
\begin{tabular}{@{}lrrrrr@{}}
\toprule
Input & $q_H$ & $q_V$ & $q_R$ & $q_L$ & $q_{\emptyset}$ \\
\midrule
H & 0.213270 & 0.005353 & 0.116738 & 0.125416 & 0.539223 \\
V & 0.006406 & 0.214433 & 0.119415 & 0.129508 & 0.530239 \\
R & 0.111548 & 0.111566 & 0.230747 & 0.005297 & 0.540842 \\
L & 0.108586 & 0.107808 & 0.006318 & 0.248648 & 0.528639 \\
D & 0.109933 & 0.110398 & 0.113396 & 0.135574 & 0.530698 \\
A & 0.110199 & 0.108956 & 0.122863 & 0.119209 & 0.538773 \\
\bottomrule
\end{tabular}
\end{minipage}\par\medskip

\noindent\begin{minipage}{\textwidth}
\centering
{\bfseries 780 nm Si P380\par}

\centering
\footnotesize
\begin{tabular}{@{}lrrrrr@{}}
\toprule
Input & $q_H$ & $q_V$ & $q_R$ & $q_L$ & $q_{\emptyset}$ \\
\midrule
H & 0.098146 & 0.006665 & 0.056365 & 0.056515 & 0.782310 \\
V & 0.006290 & 0.106701 & 0.060412 & 0.065280 & 0.761317 \\
R & 0.051032 & 0.058291 & 0.114856 & 0.002450 & 0.773371 \\
L & 0.052787 & 0.055744 & 0.002221 & 0.119116 & 0.770132 \\
D & 0.066545 & 0.072716 & 0.057073 & 0.064696 & 0.738969 \\
A & 0.069983 & 0.068522 & 0.062005 & 0.056271 & 0.743219 \\
\bottomrule
\end{tabular}
\end{minipage}\par\medskip

\noindent\begin{minipage}{\textwidth}
\centering
{\bfseries 780 nm TiO$_2$ P360\par}

\centering
\footnotesize
\begin{tabular}{@{}lrrrrr@{}}
\toprule
Input & $q_H$ & $q_V$ & $q_R$ & $q_L$ & $q_{\emptyset}$ \\
\midrule
H & 0.202871 & 0.011754 & 0.059553 & 0.048202 & 0.677619 \\
V & 0.014229 & 0.222164 & 0.035686 & 0.035660 & 0.692262 \\
R & 0.108058 & 0.114286 & 0.088817 & 0.003483 & 0.685356 \\
L & 0.107748 & 0.121073 & 0.006195 & 0.080349 & 0.684635 \\
D & 0.103121 & 0.115863 & 0.048326 & 0.040216 & 0.692473 \\
A & 0.109998 & 0.116563 & 0.045564 & 0.042516 & 0.685359 \\
\bottomrule
\end{tabular}
\end{minipage}\par\medskip

\noindent\begin{minipage}{\textwidth}
\centering
{\bfseries 2000 nm Si P900\par}

\centering
\footnotesize
\begin{tabular}{@{}lrrrrr@{}}
\toprule
Input & $q_H$ & $q_V$ & $q_R$ & $q_L$ & $q_{\emptyset}$ \\
\midrule
H & 0.200591 & 0.003779 & 0.074771 & 0.077185 & 0.643674 \\
V & 0.005485 & 0.191613 & 0.071486 & 0.065884 & 0.665532 \\
R & 0.105572 & 0.098687 & 0.143813 & 0.002620 & 0.649308 \\
L & 0.103287 & 0.093880 & 0.002314 & 0.140694 & 0.659824 \\
D & 0.100389 & 0.101871 & 0.074174 & 0.071010 & 0.652556 \\
A & 0.108455 & 0.090714 & 0.072082 & 0.072173 & 0.656575 \\
\bottomrule
\end{tabular}
\end{minipage}\par\medskip

\noindent\begin{minipage}{\textwidth}
\centering
{\bfseries 10.6 $\mu$m Si P4700\par}

\centering
\footnotesize
\begin{tabular}{@{}lrrrrr@{}}
\toprule
Input & $q_H$ & $q_V$ & $q_R$ & $q_L$ & $q_{\emptyset}$ \\
\midrule
H & 0.076795 & 0.001268 & 0.046015 & 0.047835 & 0.828087 \\
V & 0.001217 & 0.072391 & 0.051220 & 0.051913 & 0.823259 \\
R & 0.037464 & 0.035313 & 0.096252 & 0.001265 & 0.829705 \\
L & 0.040363 & 0.038510 & 0.001079 & 0.098415 & 0.821632 \\
D & 0.040218 & 0.036362 & 0.049304 & 0.049335 & 0.824781 \\
A & 0.037605 & 0.037471 & 0.047941 & 0.050426 & 0.826557 \\
\bottomrule
\end{tabular}
\end{minipage}\par\medskip

\subsection{Reconstructed measurement operators}

The coefficients below specify the four click operators through Eq.~\eqref{eq:si_operator_matrix}; the no-click operator is $\hat I-\sum_j\hat E_j$. Rounded values are printed for readability, while the security calculation uses full precision.

\begin{center}
\small
\begin{tabular}{@{}llrrrr@{}}
\toprule
Fixed receiver & Port & $t_j$ & $x_j$ & $y_j$ & $z_j$ \\
\midrule
1550 nm Si P750 & H & 0.220285 & 0.000148 & -0.002380 & 0.214362 \\
1550 nm Si P750 & V & 0.210834 & 0.001542 & -0.006740 & -0.203790 \\
1550 nm Si P750 & R & 0.189688 & 0.013998 & 0.183556 & -0.004770 \\
1550 nm Si P750 & L & 0.201740 & -0.009447 & -0.189917 & -0.005167 \\
1550 nm Si P700 & H & 0.219981 & -0.000267 & 0.002961 & 0.206864 \\
1550 nm Si P700 & V & 0.219505 & 0.001442 & 0.003758 & -0.209080 \\
1550 nm Si P700 & R & 0.236492 & -0.009466 & 0.224429 & -0.002677 \\
1550 nm Si P700 & L & 0.254551 & 0.016366 & -0.243351 & -0.004092 \\
780 nm Si P380 & H & 0.114927 & -0.003438 & -0.001755 & 0.091856 \\
780 nm Si P380 & V & 0.122879 & 0.004195 & 0.002547 & -0.100036 \\
780 nm Si P380 & R & 0.117644 & -0.004932 & 0.112635 & -0.004048 \\
780 nm Si P380 & L & 0.121443 & 0.008425 & -0.116666 & -0.008765 \\
780 nm TiO$_2$ P360 & H & 0.215342 & -0.006877 & 0.000311 & 0.188643 \\
780 nm TiO$_2$ P360 & V & 0.233901 & -0.000700 & -0.006787 & -0.210410 \\
780 nm TiO$_2$ P360 & R & 0.094714 & 0.002762 & 0.082621 & 0.023867 \\
780 nm TiO$_2$ P360 & L & 0.083475 & -0.002299 & -0.076866 & 0.012543 \\
2000 nm Si P900 & H & 0.207927 & -0.008066 & 0.002284 & 0.195106 \\
2000 nm Si P900 & V & 0.193515 & 0.011156 & 0.004807 & -0.187834 \\
2000 nm Si P900 & R & 0.146213 & 0.002092 & 0.141499 & 0.003285 \\
2000 nm Si P900 & L & 0.143189 & -0.001163 & -0.138074 & 0.011301 \\
10.6 $\mu$m Si P4700 & H & 0.077888 & 0.002613 & -0.002899 & 0.075579 \\
10.6 $\mu$m Si P4700 & V & 0.073772 & -0.001109 & -0.003197 & -0.071123 \\
10.6 $\mu$m Si P4700 & R & 0.097270 & 0.001362 & 0.095173 & -0.005205 \\
10.6 $\mu$m Si P4700 & L & 0.099730 & -0.001090 & -0.097150 & -0.004078 \\
\bottomrule
\end{tabular}
\end{center}

\FloatBarrier
\clearpage
\subsection{D/A response maps}

\begin{figure}[!htbp]
\centering
\includegraphics[width=0.83\textwidth]{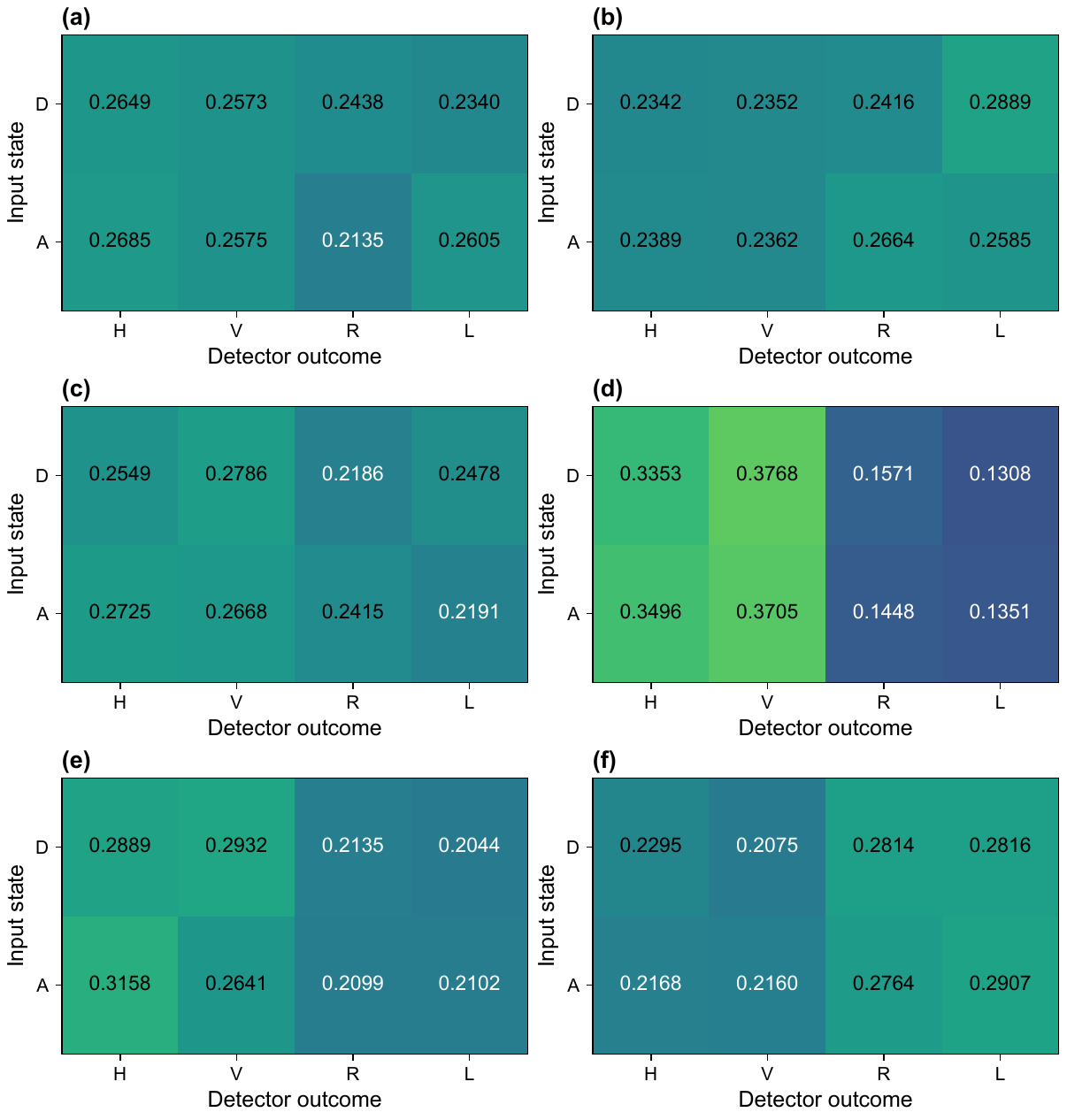}
\caption{Conditional D/A response matrices with the original fixed detector regions for \textbf{(a)} 1550~nm Si P750, \textbf{(b)} 1550~nm Si P700, \textbf{(c)} 780~nm Si P380, \textbf{(d)} 780~nm TiO$_2$ P360, \textbf{(e)} 2000~nm Si P900 and \textbf{(f)} 10.6~$\mu$m Si P4700. Rows are D and A inputs; columns are H,V,R,L detector outcomes. The printed values are normalized only by total power in the four detector regions.}
\label{fig:si_da_matrices}
\end{figure}
\FloatBarrier

\subsection{Focal fields and calibration diagnostics}

For each device below, the intensity pair shows D and A with the original physical detector boundaries used in tomography. The white crosses mark independently fitted spot centres. The growth curves are native-grid calibration diagnostics: each curve is the worst of the four ports for its input state at that square size.

\FloatBarrier
\clearpage
\subsubsection{1550~nm Si P750}
\noindent\begin{minipage}{\textwidth}\centering
\centering
\includegraphics[width=0.80\textwidth]{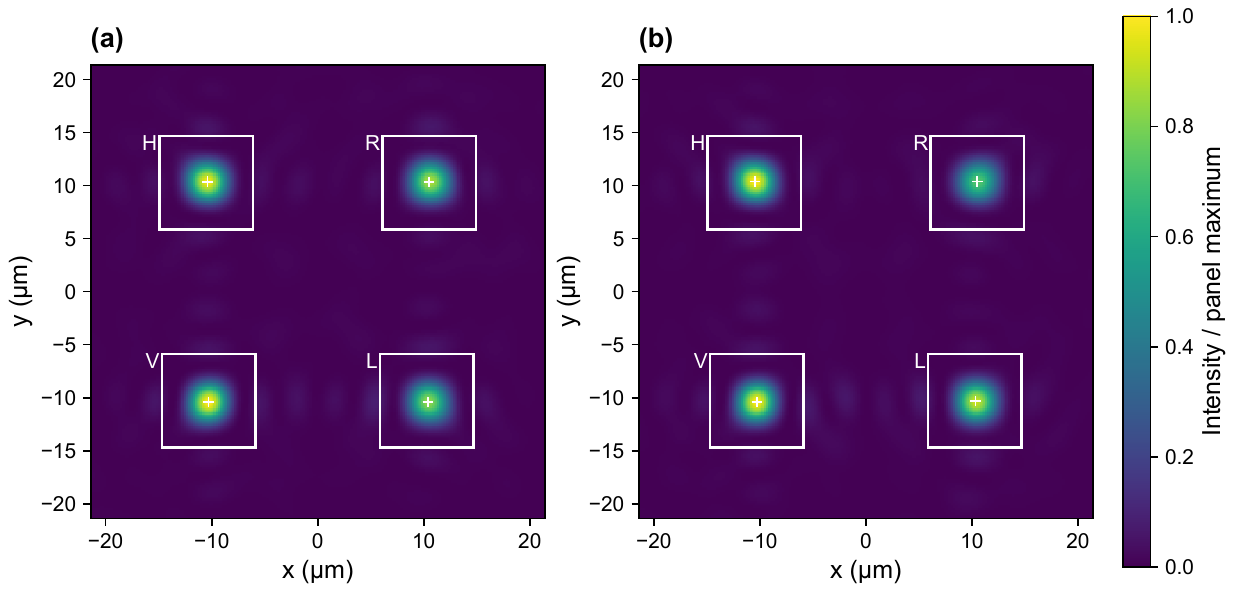}
\captionof{figure}{1550~nm Si P750 focal-plane intensities for \textbf{(a)} D and \textbf{(b)} A input. White squares are the original physical detector regions used for all six tomography inputs; crosses mark fitted centres. Each map is scaled to its own maximum for display only.}
\end{minipage}\par\vspace{5pt}
\noindent\begin{minipage}{\textwidth}\centering
\centering
\includegraphics[width=\textwidth]{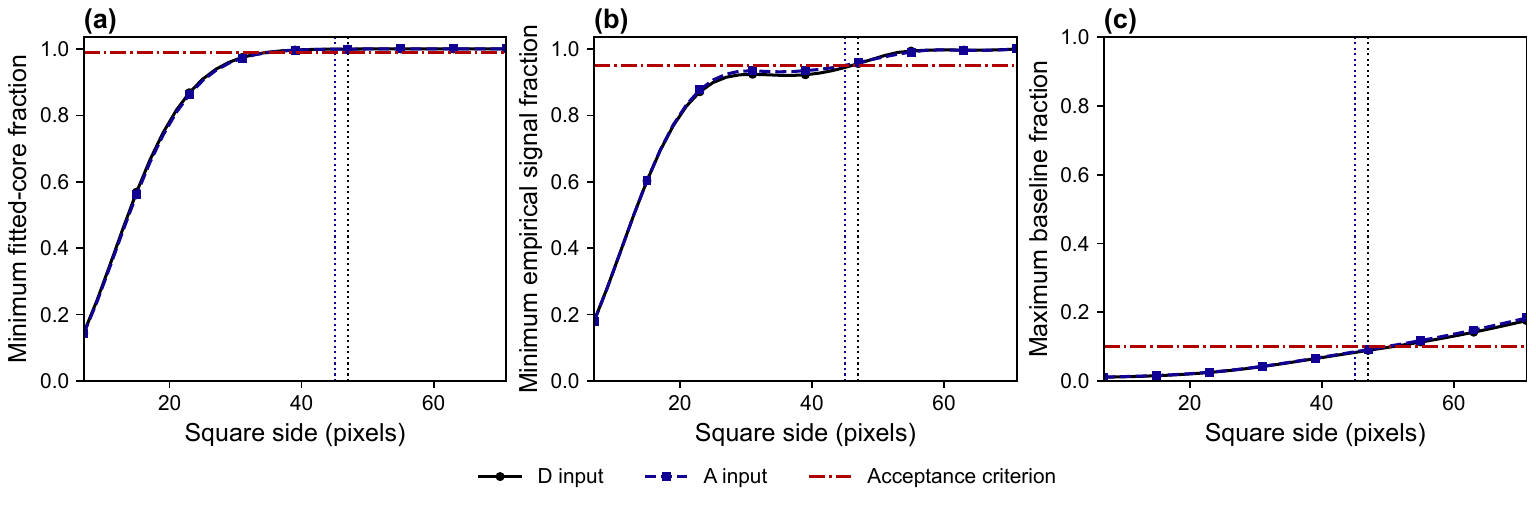}
\captionof{figure}{1550~nm Si P750 native D/A calibration: \textbf{(a)} minimum fitted-core retention, \textbf{(b)} minimum empirical net-signal fraction relative to a 71-pixel reference and \textbf{(c)} maximum estimated baseline fraction across the four ports. D uses circles and a solid curve; A uses squares and a dashed curve. Horizontal dash-dot lines mark the acceptance criteria; vertical dotted lines mark the selected common square sizes for each input.}
\end{minipage}\par\vspace{5pt}
\noindent\begin{minipage}{\textwidth}
\captionof{table}{1550~nm Si P750 D/A Gaussian-fit parameters. $B/A$ is baseline divided by fitted peak amplitude. Coordinates and core widths are in the native physical frame.}

\centering
\footnotesize
\begin{tabular}{@{}llrrrrr@{}}
\toprule
Input & Port & $x_0$ ($\mu$m) & $y_0$ ($\mu$m) & $\sigma_x$ ($\mu$m) & $\sigma_y$ ($\mu$m) & $B/A$ \\
\midrule
D & H & -10.4044 & 10.3713 & 1.2816 & 1.2670 & 0.00730 \\
D & V & -10.3036 & -10.4020 & 1.2169 & 1.2614 & 0.00653 \\
D & R & 10.5182 & 10.3633 & 1.2967 & 1.2852 & 0.00860 \\
D & L & 10.3844 & -10.3941 & 1.2808 & 1.2909 & 0.00998 \\
A & H & -10.4454 & 10.3992 & 1.2725 & 1.2528 & 0.00713 \\
A & V & -10.2931 & -10.3979 & 1.2084 & 1.2563 & 0.00762 \\
A & R & 10.5079 & 10.3823 & 1.3052 & 1.3043 & 0.01049 \\
A & L & 10.3410 & -10.3396 & 1.2879 & 1.3110 & 0.00809 \\
\bottomrule
\end{tabular}

\end{minipage}\par

\FloatBarrier
\clearpage
\subsubsection{1550~nm Si P700}
\noindent\begin{minipage}{\textwidth}\centering
\centering
\includegraphics[width=0.80\textwidth]{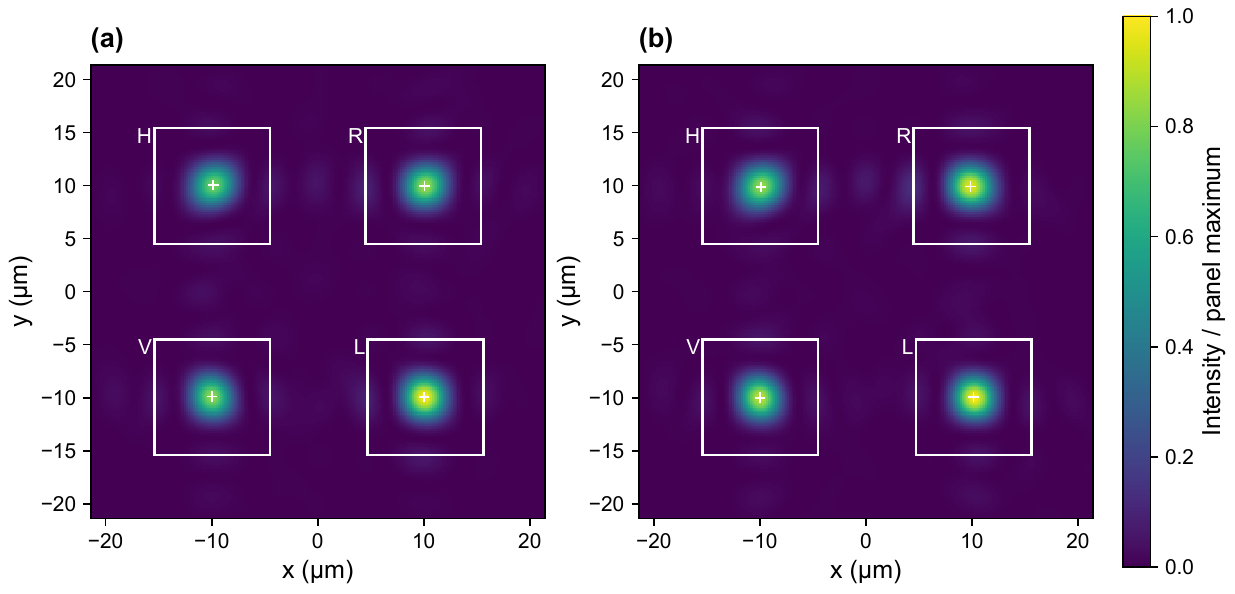}
\captionof{figure}{1550~nm Si P700 focal-plane intensities for \textbf{(a)} D and \textbf{(b)} A input. White squares are the original physical detector regions used for all six tomography inputs; crosses mark fitted centres. Each map is scaled to its own maximum for display only.}
\end{minipage}\par\vspace{5pt}
\noindent\begin{minipage}{\textwidth}\centering
\centering
\includegraphics[width=\textwidth]{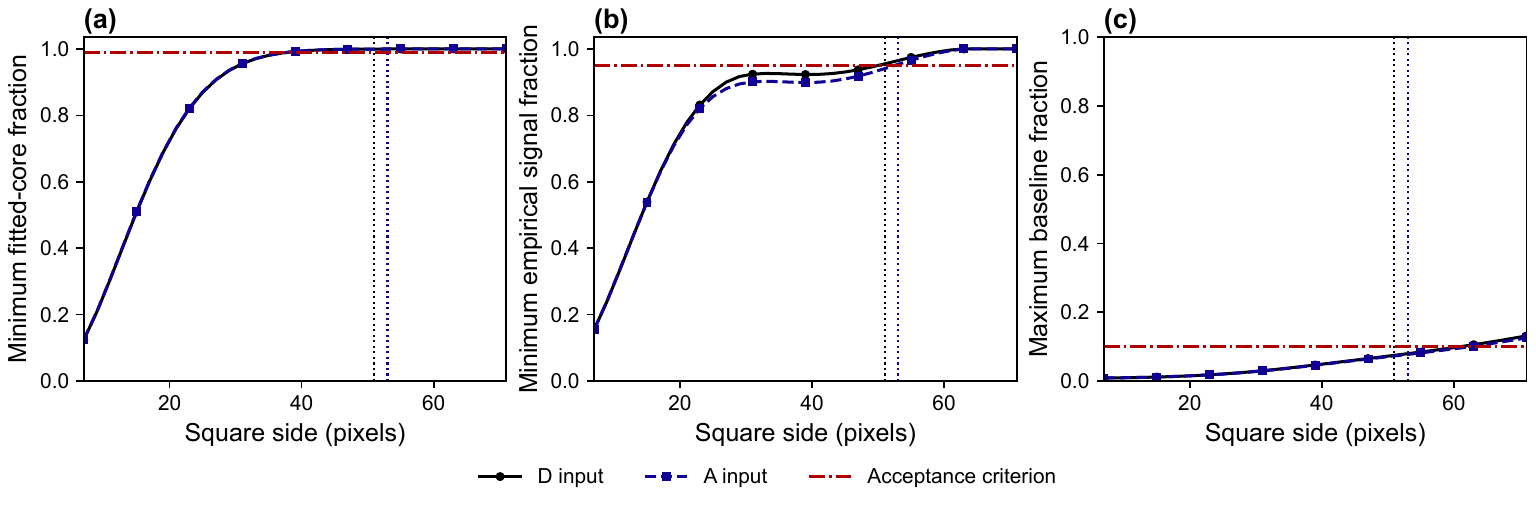}
\captionof{figure}{1550~nm Si P700 native D/A calibration: \textbf{(a)} minimum fitted-core retention, \textbf{(b)} minimum empirical net-signal fraction relative to a 71-pixel reference and \textbf{(c)} maximum estimated baseline fraction across the four ports. D uses circles and a solid curve; A uses squares and a dashed curve. Horizontal dash-dot lines mark the acceptance criteria; vertical dotted lines mark the selected common square sizes for each input.}
\end{minipage}\par\vspace{5pt}
\noindent\begin{minipage}{\textwidth}
\captionof{table}{1550~nm Si P700 D/A Gaussian-fit parameters. $B/A$ is baseline divided by fitted peak amplitude. Coordinates and core widths are in the native physical frame.}

\centering
\footnotesize
\begin{tabular}{@{}llrrrrr@{}}
\toprule
Input & Port & $x_0$ ($\mu$m) & $y_0$ ($\mu$m) & $\sigma_x$ ($\mu$m) & $\sigma_y$ ($\mu$m) & $B/A$ \\
\midrule
D & H & -9.8466 & 10.0351 & 1.4251 & 1.3877 & 0.00510 \\
D & V & -9.9894 & -9.8918 & 1.3157 & 1.3955 & 0.00733 \\
D & R & 10.0642 & 9.9579 & 1.3517 & 1.3508 & 0.00774 \\
D & L & 9.9985 & -9.9176 & 1.3643 & 1.3963 & 0.00671 \\
A & H & -9.8861 & 9.8883 & 1.4193 & 1.3869 & 0.00764 \\
A & V & -9.9887 & -10.0023 & 1.3066 & 1.3925 & 0.00616 \\
A & R & 9.8533 & 9.9245 & 1.3992 & 1.3608 & 0.00790 \\
A & L & 10.1476 & -9.9541 & 1.3211 & 1.3668 & 0.00700 \\
\bottomrule
\end{tabular}

\end{minipage}\par

\FloatBarrier
\clearpage
\subsubsection{780~nm Si P380}
\noindent\begin{minipage}{\textwidth}\centering
\centering
\includegraphics[width=0.80\textwidth]{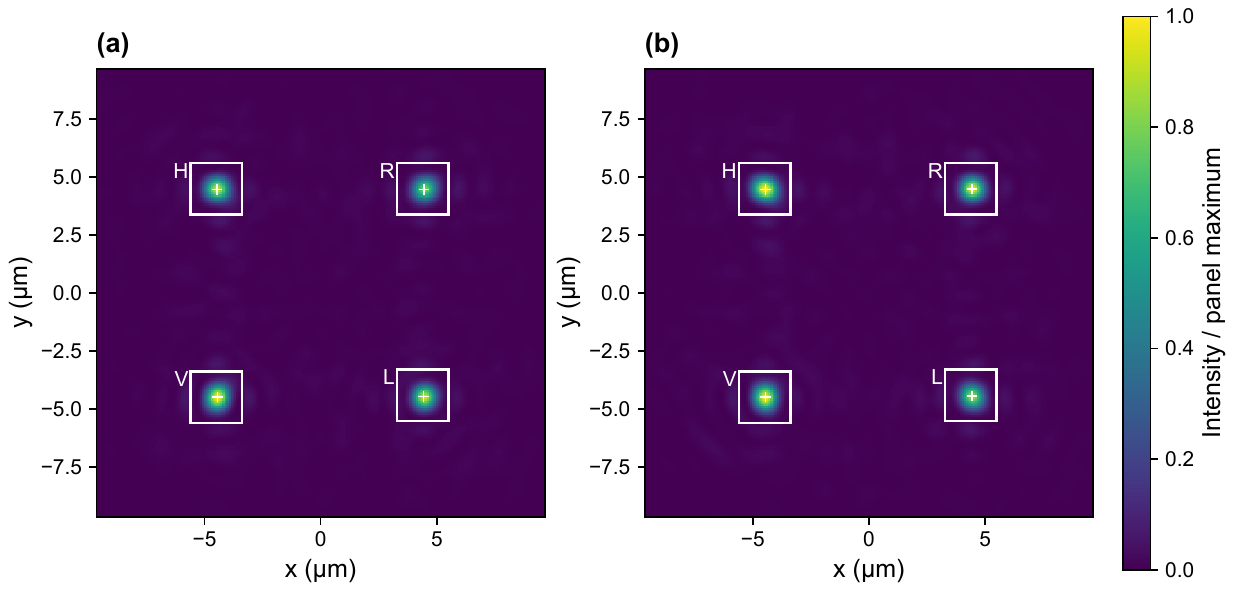}
\captionof{figure}{780~nm Si P380 focal-plane intensities for \textbf{(a)} D and \textbf{(b)} A input. White squares are the original physical detector regions used for all six tomography inputs; crosses mark fitted centres. Each map is scaled to its own maximum for display only.}
\end{minipage}\par\vspace{5pt}
\noindent\begin{minipage}{\textwidth}\centering
\centering
\includegraphics[width=\textwidth]{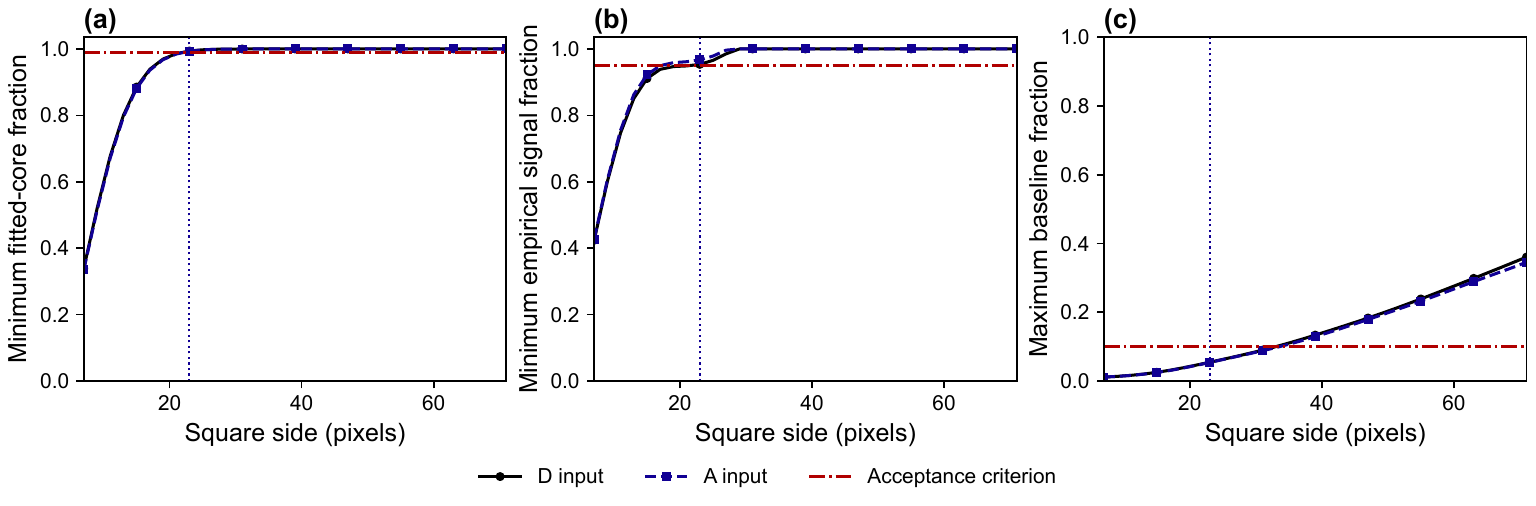}
\captionof{figure}{780~nm Si P380 native D/A calibration: \textbf{(a)} minimum fitted-core retention, \textbf{(b)} minimum empirical net-signal fraction relative to a 71-pixel reference and \textbf{(c)} maximum estimated baseline fraction across the four ports. D uses circles and a solid curve; A uses squares and a dashed curve. Horizontal dash-dot lines mark the acceptance criteria; vertical dotted lines mark the selected common square sizes for each input.}
\end{minipage}\par\vspace{5pt}
\noindent\begin{minipage}{\textwidth}
\captionof{table}{780~nm Si P380 D/A Gaussian-fit parameters. $B/A$ is baseline divided by fitted peak amplitude. Coordinates and core widths are in the native physical frame.}

\centering
\footnotesize
\begin{tabular}{@{}llrrrrr@{}}
\toprule
Input & Port & $x_0$ ($\mu$m) & $y_0$ ($\mu$m) & $\sigma_x$ ($\mu$m) & $\sigma_y$ ($\mu$m) & $B/A$ \\
\midrule
D & H & -4.4509 & 4.4646 & 0.3699 & 0.3450 & 0.00846 \\
D & V & -4.4411 & -4.4612 & 0.3340 & 0.3728 & 0.00803 \\
D & R & 4.4392 & 4.4636 & 0.3587 & 0.3498 & 0.00853 \\
D & L & 4.4303 & -4.4549 & 0.3427 & 0.3470 & 0.00864 \\
A & H & -4.4507 & 4.4567 & 0.3751 & 0.3455 & 0.00792 \\
A & V & -4.4475 & -4.4619 & 0.3377 & 0.3784 & 0.00864 \\
A & R & 4.4289 & 4.4768 & 0.3515 & 0.3394 & 0.00801 \\
A & L & 4.4273 & -4.4335 & 0.3478 & 0.3546 & 0.00877 \\
\bottomrule
\end{tabular}

\end{minipage}\par

\FloatBarrier
\clearpage
\subsubsection{780~nm TiO$_2$ P360}
\noindent\begin{minipage}{\textwidth}\centering
\centering
\includegraphics[width=0.80\textwidth]{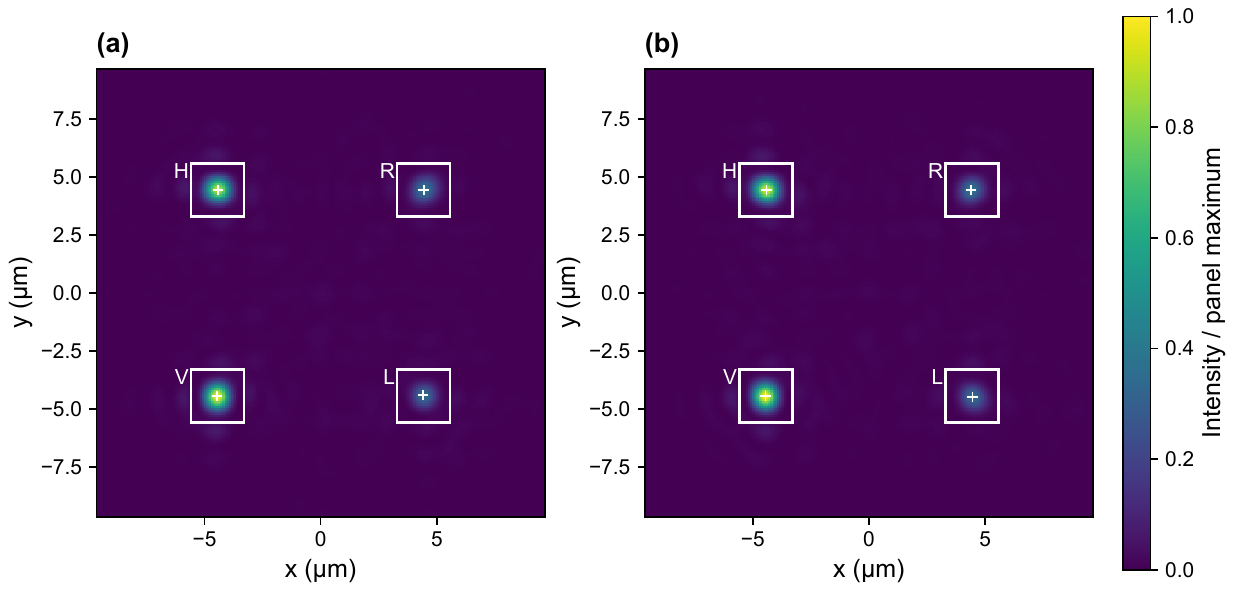}
\captionof{figure}{780~nm TiO$_2$ P360 focal-plane intensities for \textbf{(a)} D and \textbf{(b)} A input. White squares are the original physical detector regions used for all six tomography inputs; crosses mark fitted centres. Each map is scaled to its own maximum for display only.}
\end{minipage}\par\vspace{5pt}
\noindent\begin{minipage}{\textwidth}\centering
\centering
\includegraphics[width=\textwidth]{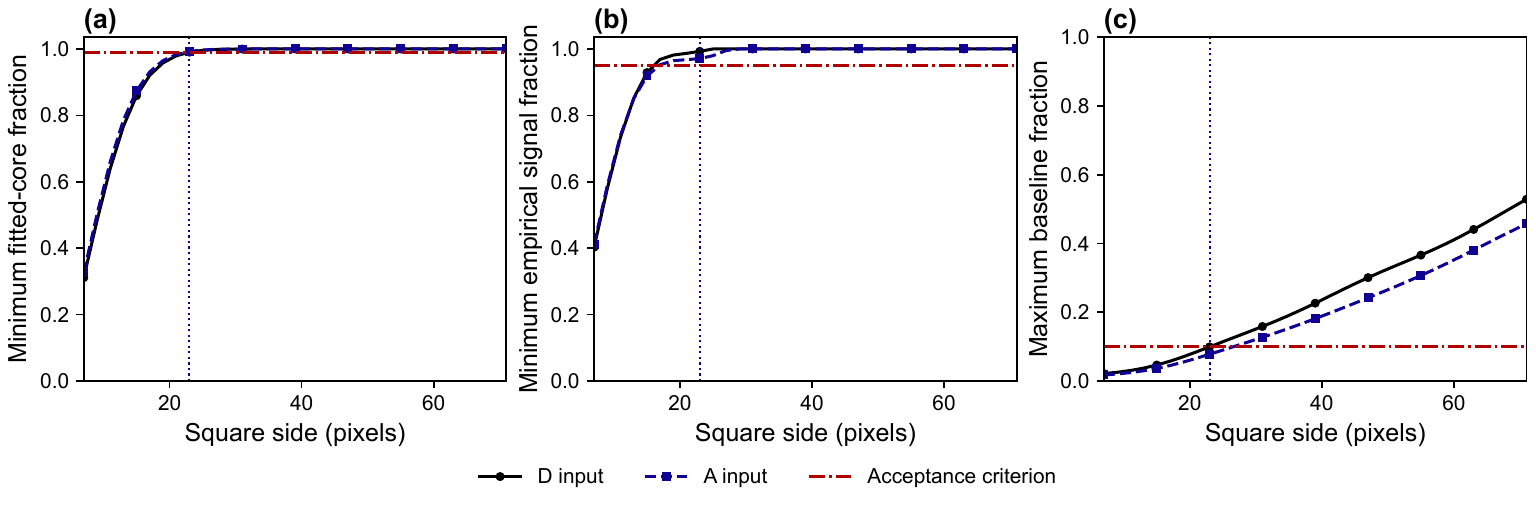}
\captionof{figure}{780~nm TiO$_2$ P360 native D/A calibration: \textbf{(a)} minimum fitted-core retention, \textbf{(b)} minimum empirical net-signal fraction relative to a 71-pixel reference and \textbf{(c)} maximum estimated baseline fraction across the four ports. D uses circles and a solid curve; A uses squares and a dashed curve. Horizontal dash-dot lines mark the acceptance criteria; vertical dotted lines mark the selected common square sizes for each input.}
\end{minipage}\par\vspace{5pt}
\noindent\begin{minipage}{\textwidth}
\captionof{table}{780~nm TiO$_2$ P360 D/A Gaussian-fit parameters. $B/A$ is baseline divided by fitted peak amplitude. Coordinates and core widths are in the native physical frame.}

\centering
\footnotesize
\begin{tabular}{@{}llrrrrr@{}}
\toprule
Input & Port & $x_0$ ($\mu$m) & $y_0$ ($\mu$m) & $\sigma_x$ ($\mu$m) & $\sigma_y$ ($\mu$m) & $B/A$ \\
\midrule
D & H & -4.4244 & 4.4428 & 0.3717 & 0.3504 & 0.00723 \\
D & V & -4.4557 & -4.4287 & 0.3511 & 0.3751 & 0.00630 \\
D & R & 4.4355 & 4.4456 & 0.3796 & 0.3738 & 0.01166 \\
D & L & 4.4142 & -4.3980 & 0.3519 & 0.3450 & 0.01656 \\
A & H & -4.4074 & 4.4263 & 0.3740 & 0.3540 & 0.00724 \\
A & V & -4.4472 & -4.4195 & 0.3506 & 0.3760 & 0.00638 \\
A & R & 4.3854 & 4.4265 & 0.3625 & 0.3572 & 0.01367 \\
A & L & 4.4627 & -4.4621 & 0.3679 & 0.3644 & 0.01323 \\
\bottomrule
\end{tabular}

\end{minipage}\par

\FloatBarrier
\clearpage
\subsubsection{2000~nm Si P900}
\noindent\begin{minipage}{\textwidth}\centering
\centering
\includegraphics[width=0.80\textwidth]{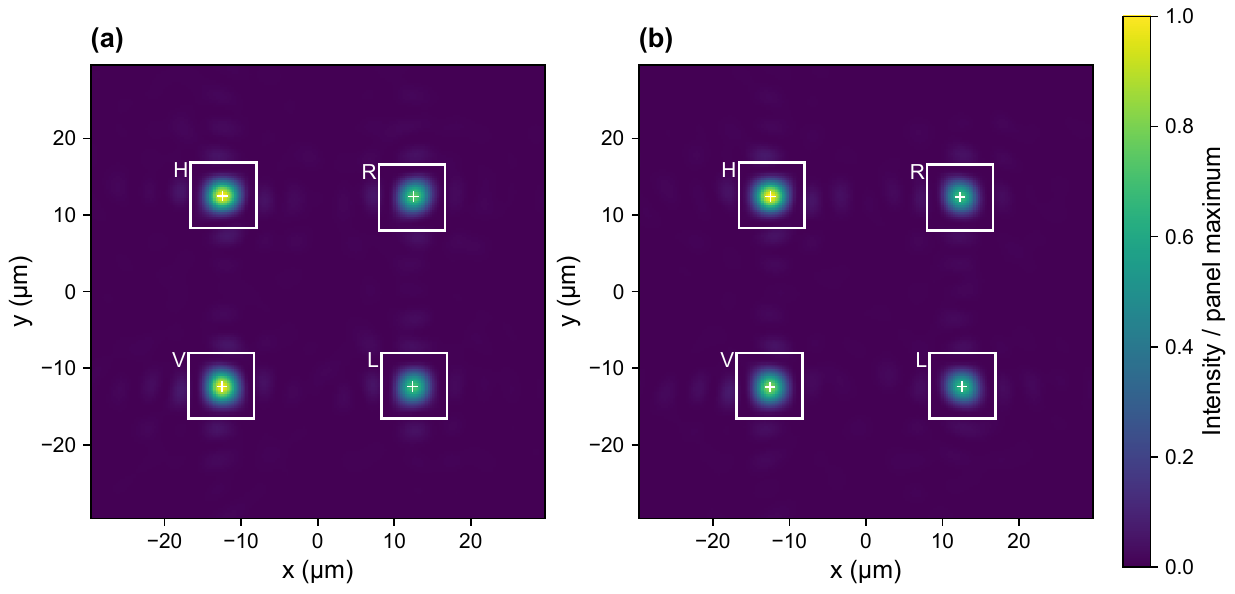}
\captionof{figure}{2000~nm Si P900 focal-plane intensities for \textbf{(a)} D and \textbf{(b)} A input. White squares are the original physical detector regions used for all six tomography inputs; crosses mark fitted centres. Each map is scaled to its own maximum for display only.}
\end{minipage}\par\vspace{5pt}
\noindent\begin{minipage}{\textwidth}\centering
\centering
\includegraphics[width=\textwidth]{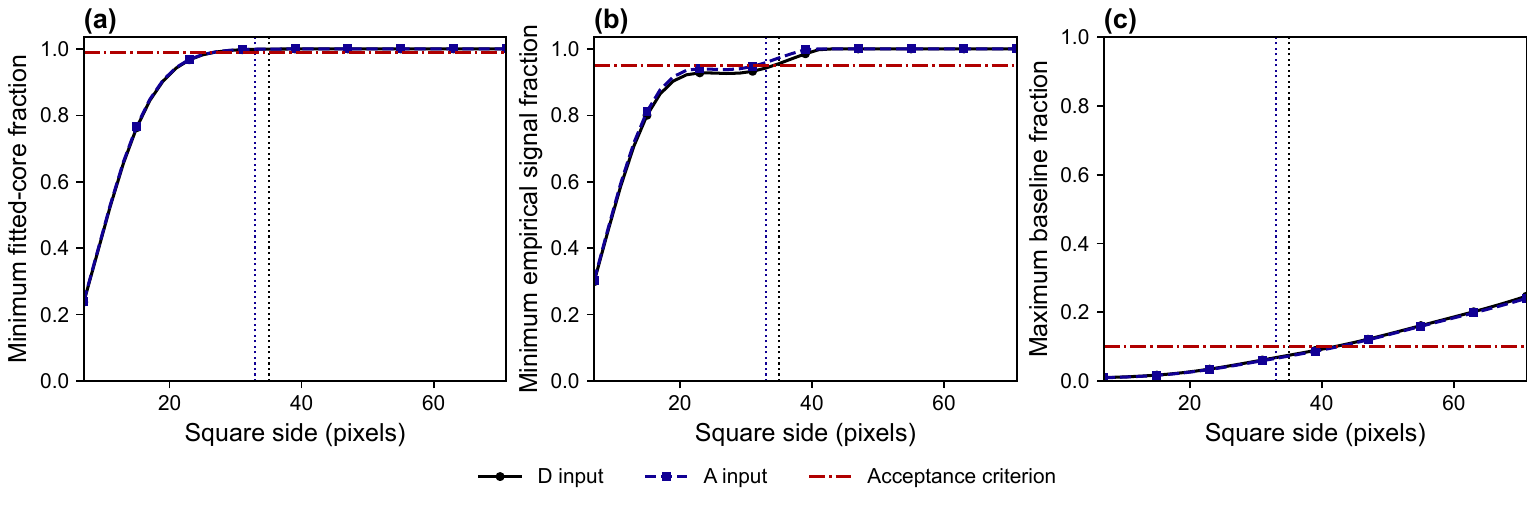}
\captionof{figure}{2000~nm Si P900 native D/A calibration: \textbf{(a)} minimum fitted-core retention, \textbf{(b)} minimum empirical net-signal fraction relative to a 71-pixel reference and \textbf{(c)} maximum estimated baseline fraction across the four ports. D uses circles and a solid curve; A uses squares and a dashed curve. Horizontal dash-dot lines mark the acceptance criteria; vertical dotted lines mark the selected common square sizes for each input.}
\end{minipage}\par\vspace{5pt}
\noindent\begin{minipage}{\textwidth}
\captionof{table}{2000~nm Si P900 D/A Gaussian-fit parameters. $B/A$ is baseline divided by fitted peak amplitude. Coordinates and core widths are in the native physical frame.}

\centering
\footnotesize
\begin{tabular}{@{}llrrrrr@{}}
\toprule
Input & Port & $x_0$ ($\mu$m) & $y_0$ ($\mu$m) & $\sigma_x$ ($\mu$m) & $\sigma_y$ ($\mu$m) & $B/A$ \\
\midrule
D & H & -12.4461 & 12.5115 & 1.3888 & 1.3002 & 0.00611 \\
D & V & -12.4924 & -12.3911 & 1.3265 & 1.3910 & 0.00703 \\
D & R & 12.4970 & 12.4487 & 1.3230 & 1.3275 & 0.00827 \\
D & L & 12.3993 & -12.4109 & 1.3248 & 1.3539 & 0.00723 \\
A & H & -12.4544 & 12.4487 & 1.3760 & 1.3027 & 0.00584 \\
A & V & -12.5080 & -12.4708 & 1.3216 & 1.3790 & 0.00673 \\
A & R & 12.2908 & 12.3921 & 1.2892 & 1.3205 & 0.00723 \\
A & L & 12.5263 & -12.4050 & 1.3417 & 1.2986 & 0.00790 \\
\bottomrule
\end{tabular}

\end{minipage}\par

\FloatBarrier
\clearpage
\subsubsection{10.6~$\mu$m Si P4700}
\noindent\begin{minipage}{\textwidth}\centering
\centering
\includegraphics[width=0.80\textwidth]{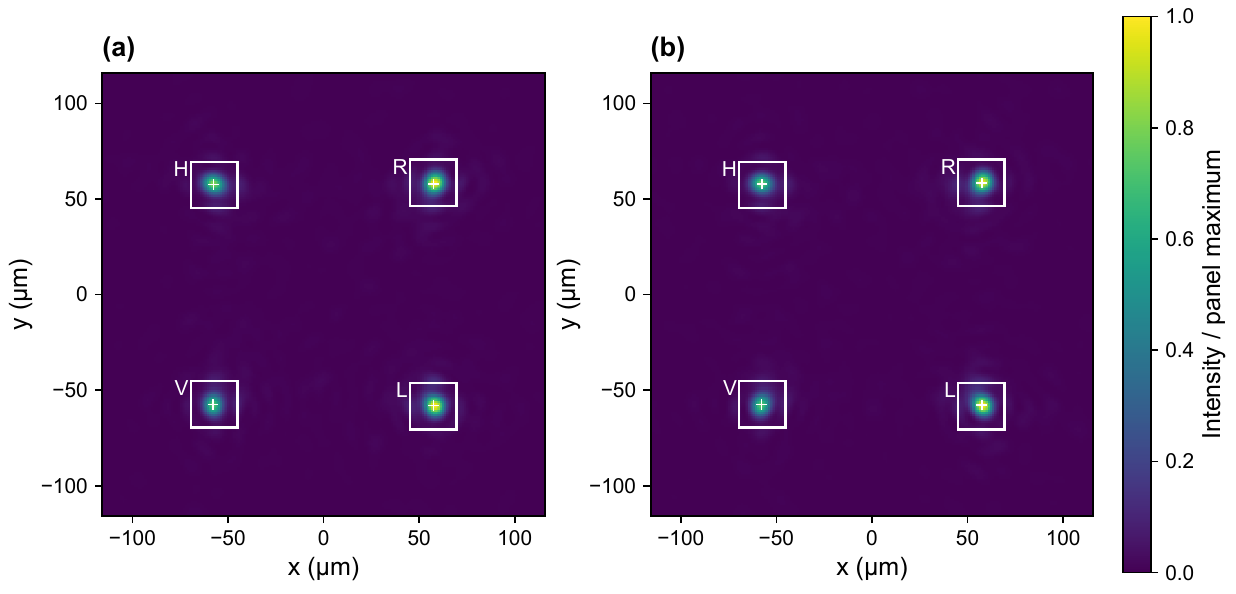}
\captionof{figure}{10.6~$\mu$m Si P4700 focal-plane intensities for \textbf{(a)} D and \textbf{(b)} A input. White squares are the original physical detector regions used for all six tomography inputs; crosses mark fitted centres. Each map is scaled to its own maximum for display only.}
\end{minipage}\par\vspace{5pt}
\noindent\begin{minipage}{\textwidth}\centering
\centering
\includegraphics[width=\textwidth]{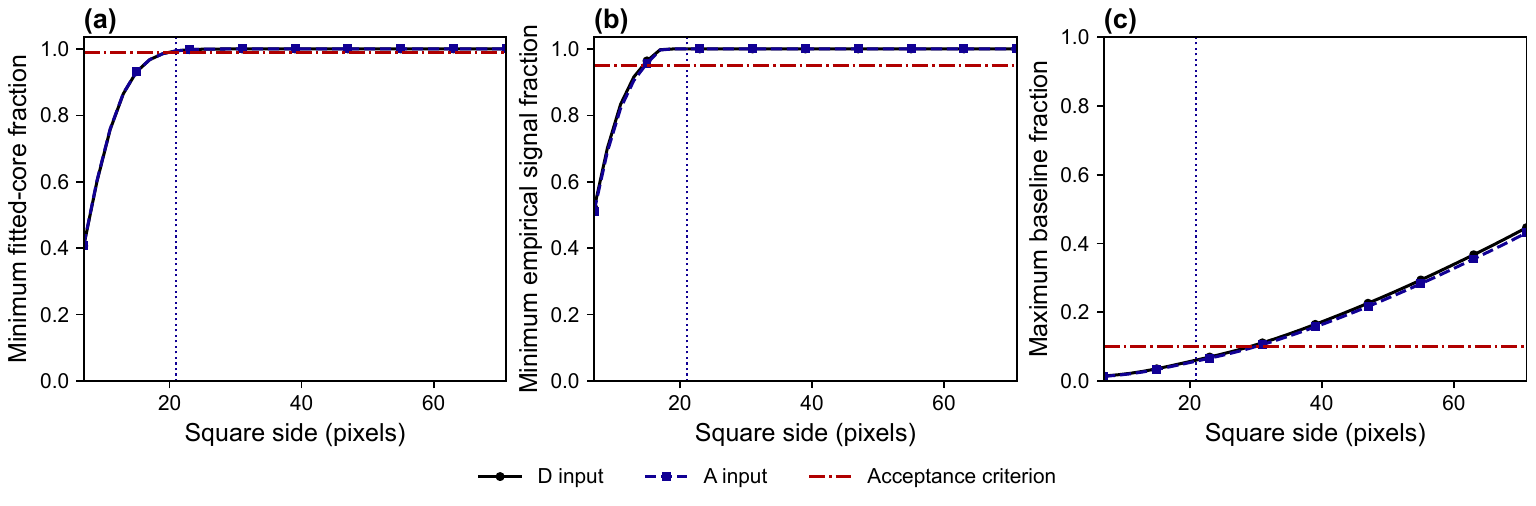}
\captionof{figure}{10.6~$\mu$m Si P4700 native D/A calibration: \textbf{(a)} minimum fitted-core retention, \textbf{(b)} minimum empirical net-signal fraction relative to a 71-pixel reference and \textbf{(c)} maximum estimated baseline fraction across the four ports. D uses circles and a solid curve; A uses squares and a dashed curve. Horizontal dash-dot lines mark the acceptance criteria; vertical dotted lines mark the selected common square sizes for each input.}
\end{minipage}\par\vspace{5pt}
\noindent\begin{minipage}{\textwidth}
\captionof{table}{10.6~$\mu$m Si P4700 D/A Gaussian-fit parameters. $B/A$ is baseline divided by fitted peak amplitude. Coordinates and core widths are in the native physical frame.}

\centering
\footnotesize
\begin{tabular}{@{}llrrrrr@{}}
\toprule
Input & Port & $x_0$ ($\mu$m) & $y_0$ ($\mu$m) & $\sigma_x$ ($\mu$m) & $\sigma_y$ ($\mu$m) & $B/A$ \\
\midrule
D & H & -57.5361 & 57.4979 & 4.0894 & 3.5344 & 0.00907 \\
D & V & -57.8528 & -57.5646 & 3.4494 & 4.0986 & 0.01039 \\
D & R & 57.5171 & 57.9512 & 3.7326 & 3.8240 & 0.00945 \\
D & L & 57.5106 & -58.0423 & 3.7322 & 3.7917 & 0.00933 \\
A & H & -57.6062 & 57.6923 & 3.9996 & 3.4409 & 0.00957 \\
A & V & -57.8161 & -57.5732 & 3.4872 & 4.1095 & 0.00976 \\
A & R & 57.3977 & 58.3404 & 3.7388 & 3.7350 & 0.00941 \\
A & L & 57.3921 & -57.7597 & 3.6726 & 3.7975 & 0.00935 \\
\bottomrule
\end{tabular}

\end{minipage}\par
\FloatBarrier

\FloatBarrier

\end{document}